\documentclass[a4paper,aps,prb,twocolumn]{revtex4}
\usepackage{amsmath}
\usepackage{amssymb}
\usepackage{graphicx}
\usepackage{overpic}
\usepackage{subfigure}

\def \dd{\mathrm{d}}
\DeclareMathOperator{\re}{Re}
\DeclareMathOperator{\im}{Im}
\DeclareMathOperator{\sgn}{sgn}

\def \j{\vec j}
\def \q{\vec q}
\def \r{\vec r}
\def \R{\vec R}
\def \A{\vec A}
\def \a{\vec a}

\def \k{\vec k}

\def \O{\Omega}
\def \o{\omega}

\def \dr{\delta\rho}
\def \la{\langle}
\def \ra{\rangle}
\def \be{\begin{equation}}
\def \ee{\end{equation}}
\def \ba{\begin{align*}}
\def \ea{\end{align*}}
\def \ben{\begin{eqnarray}}
\def \een{\end{eqnarray}}
\def \mD{\mathcal D}
\def \mZ{\mathcal Z}

\begin{document}
\title{Theoretical Analysis of Drag Resistance \\ in Amorphous Thin Films Exhibiting Superconductor-Insulator-Transition}
\author{Yue Zou${}^1$, Gil Refael${}^1$, and Jongsoo Yoon${}^2$}
\affiliation{${}^1$Department of Physics, California Institute of
Technology, Pasadena, California 91125, USA \\ ${}^2$Department of Physics, University of Virginia, Charlottesville, Virginia 22903, USA}
\date{\today}

\begin{abstract}
The magnetical field tuned superconductor-insulator transition in amorphous thin
films, e.g., Ta and InO, exhibits a range of yet unexplained curious phenomena, such as a putative
low-resistance metallic phase intervening the superconducting and the insulating phase, and a huge peak in the
magnetoresistance at large magnetic field. Qualitatively, the phenomena can be explained equally well within
several significantly different pictures, particularly the condensation of quantum
vortex liquid, and the percolation of superconducting islands embedded in normal region. Recently, we proposed and analyzed a new measurement in Ref. \onlinecite{drag} that should be able
to decisively point to the correct picture: a
drag resistance measurement in an amorphous thin-film bilayer setup. Neglecting interlayer tunneling,
we found that the drag resistance
within the vortex paradigm has opposite sign and is orders of magnitude
larger than that in competing paradigms. For example, two identical films as in Ref. \onlinecite{shahar2004} with $25$nm layer separation at $0.07$K would produce a drag resistance $\sim10^{-4}\Omega$ according the vortex theory, but only $\sim10^{-12}\Omega$ for the percolation theory. We provide details of our theoretical analysis of the drag resistance within both paradigms, and report some new results as well.
\end{abstract}
\maketitle

\section{Introduction}
Amorphous thin film superconductors exhibit a variety of fascinating
quantum phenomena, due to the importance of fluctuation and disorder
in two dimensions. Early
theoretical\cite{FisherLee,Fisher1990,WenZee,FisherGrinsteinGirvin,Cha1991,Wallin1994}
and experimental\cite{Haviland1989,Hebard1990,Hebard1992,Valles1992,Liu1993,Hsu1995,Valles1994,
Yazdani1995,Hsu1998,Goldman1998} work focus on the quantum
superconductor-insulator-transition (SIT) in these materials. As one
increases the perpendicular magnetic field or decreases the film
thickness, the film changes from superconducting to insulating. An
appealing theoretical picture of the SIT is that the amplitude of
the superconducting order parameter remains finite across the
transition, and the transition is driven by phase fluctuations, which
can be viewed as the condensation of vortices. Therefore the insulator is described as a
vortex superfluid, and the transition point is nearly self dual: it
could be described either as the condensation of Cooper pairs, or of
vortices. This Cooper-pair - vortex duality also suggests that the
critical resistance at the transition should be
$R_{\square}=h/4e^2=6.5k\Omega$, which is consistent with observations
on strongly disordered samples \cite{Kapitulnik2007}. A variety of
other experiments shows a transition with a critical resistance of the
same order as  $R_Q=h/4e^2$. 

\begin{figure}[h]
\centering
\includegraphics[scale=0.42]{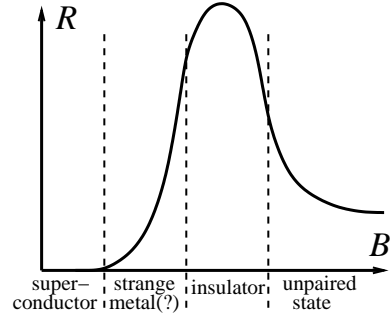}
\caption{A typical magnetoresistance curve of amorphous thin film superconductors. As the magnetic field $B$ increases, the superconducting phase is destroyed, and a possible metallic phase emerges. After which the system enters an insulating phase, where the magnetoresistance reaches its peak. The resistance drops down and approaches normal state value as $B$ is further increased.}\label{fig:pd}
\end{figure}

In recent years, experiments on these amorphous thin films have
revealed more surprising results, mainly in transitions tuned by
normal magnetic field. One of these raises the possibility that a
metallic phase intervenes between the superconducting and the insulating
phases\cite{Kapitulnik1996,Kapitulnik1999,Kapitulnik2001,shahar2004,Kapitulnik2005,Yoon2005,Yoon2006,Yoon2010}. Near
the "SIT critical point", as temperature is lowered below $\sim 100$mK, the
resistance curve starts to level off, indicating the existence of a
novel metallic phase, with a distinct nonlinear $I-V$ characteristics
at least in Ta films that are interpreted as a consequence of vortex dynamics \cite{Yoon2005}. Another interesting experimental finding is the
nonmonotonic behavior of the
magnetoresistance\cite{shahar2004,Baturina2004,Kapitulnik2005,review}. As one
increases magnetic field further from the "SIT point", the resistance
climbs up quickly to very large value in InO and TiN films, before it plummeting back to
the normal state resistance, as shown in FIG. \ref{fig:pd}. In Ta
and MoGe films, as well as some InO films, the resistance peak is
not as large, but is still apparent \cite{Kapitulnik1996,Kapitulnik1999,Kapitulnik2001, Kapitulnik2005,Yoon2005,Yoon2006}.

Two competing paradigms may account for these phenomena. On the one
hand, within the quantum vortex pictures
\cite{Fisher1990,Feigelman1993,RotonFermiLiquid,vortexmetal}, the insulating phase at the peak of the magnetoresistance implies the condensation of quantum vortices, and the high field negative
magnetoresistance indicates the gradual depairing of Cooper pairs and the appearance of a finite electronic density of
states at the Fermi level. The intervening metallic phase is described
as a delocalzed but yet uncondensed diffusive vortex liquid as
described in Ref. \onlinecite{vortexmetal}. In this picture disorder
and charging effects are most important on length scales smaller or of order
$\xi$ (the superconducting coherence length, typically of order $10nm$).  On the other hand, the percolation
paradigm\cite{Shimshoni1998,Trivedi2001,meir,Dubi2007,Spivak2008}
describes the amorphous film as a mixture of superconductor and normal
or insulating puddles, with disorder playing a role at scales larger
than $\xi$. Particularly germane is the picture in
Ref. \onlinecite{Dubi2007} which phenomenologically captures both a metallic phase as
well as the strongly insulating phase by assuming superconducting
islands exhibit a Coulomb blockade for electrons. This way the peak in
the magnetoresistance arises from electron transport though the percolating normal regions
consisting of narrow conduction channels. Yet a third theory tries to account for the low field
superconductor-metal transition using a phase glass model
\cite{Phillips2002,WuPhillips} (see, however, Ref. \onlinecite{Ikeda2007} which argues against these
results), but does not address the full
magnetoresistance curve. Qualitatively, both
paradigms above are consistent with magnetoresistance observations, and
recent tilted field\cite{Shahar2006}, AC conductance\cite{Armitage}, Nernst
effect\cite{Spathis2007}, and Scanning Tunneling
Spectroscopic\cite{Baturina2008} measurements cannot
distinguish between them. Particularly intriguing is the origin of the
metallic phase - is it vortex driven or does it occur due to
electronic conduction channels dominating transport through the film?

\begin{figure}[h]
\centering
\includegraphics[scale=0.45]{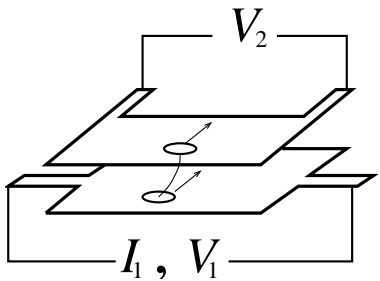}
\caption{Our proposed bilayer setup for the drag resistance measurement. A current bias $I_1$ is applied in one layer, and a voltage $V_2$ is measured in the other layer. The drag resistance $R_D$ is defined as $R_D=V_2/I_1$.}\label{setup}
\end{figure}

Given the similarity in the predictions of the distinct
vortex-condensation and percolation paradigms, an experiment that
distinguishes between them would be highly desirable. We propose that a thin film "Giaever transformer"\cite{Giaever1}
experiment (FIG. \ref{setup}) can qualitatively distinguish between these two paradigms. The original design of a
Giaever transformer consists of two type-II superconductors
separated by an insulating layer in perpendicular magnetic fields. A current in one layer moves the
vortex lattice in the entire junction, yielding the same DC voltage in
both layers. Determining the drag resistance $R_D=V_2/I_1$ in a similar bilayer structure of two amorphous
superconducting thin films should qualitatively distinguish between the two
paradigms (see also Refs. \onlinecite{Michaeli2006, Kapitulnik2002}):
within the vortex paradigm,
vortices in one layer drag
the vortices in the other, but within the percolation picture, the
drag resistance is solely due to
interlayer "Coulomb drag", as studied in semiconductor
 heterostructures \cite{Gramila}.

The first qualitative difference between vortex drag and Coulomb drag
is the sign of the drag voltage $V_2$. Denoting the voltage drop in
the driving layer as $V_1$, it is easy to see that $V_1$ and $V_2$
have the same sign if they are produced by vortex motion, because
vortices in the two layers move in the same direction transverse to
the current bias $I_1$. (We note in passing that if the second layer is in a closed circuit, the vortex drag would induce a current in the opposite
direction in the secondary layer, since no outside voltage source
balances the EMF produced by the vortex motion.) On the other hand, $V_1$ and $V_2$ would have
opposite signs if they are due to electron Coulomb drag, because $V_2$
has to balance the drag force to ensure the open circuit condition in
the second layer. In other words, Coulomb drag would try to produce
current in the same direction in the primary and secondary layer.  

More importantly, we have found that in the vortex scenario, the drag
resistance is expected to be several orders of magnitude larger than
that in other models. Partially this is expected because in these films, the
sheet carrier density $\sim 10^{16}$cm$^{-2}$ is much larger than the
vortex density $\sim B/\Phi_0\sim 10^{10}$cm$^{-2}$, and the drag
effect is typically smaller for larger densities. For example, two
identical films as in FIG. 2(b) of Ref. \onlinecite{shahar2004} with
$25$nm center-to-center layer separation at $0.07$K would produce a
drag resistance $\sim10^{-4}\Omega$ according the vortex theory (see
FIG. \ref{vortex}), but only $\sim10^{-12}\Omega$ for the percolation
theory (see FIG. \ref{percolation}). But as we shall show below, the
large vortex drag effect is also a consequence of the extremely high
magneto-resistance slope, which has different implications for the
vortex condensation and percolation pictures. The strength of the
thin-film Giaever tranformer experiment would therefore be in the
transition region where the metallic phase transforms into the
insulating phase, and the magneto-resistance is at a maximum.

We believe that these qualitative
differences between the drags in the two paradigms are quite general
for each paradigm, and does not depend the various microscopic
assumptions made in various flavors of these phenomenological
pictures.  We will support
these claims by analyzing the drag resistance between two identical thin films within a representative
theoretical framework in the vortex \cite{vortexmetal} and percolation
paradigms \cite{meir}. We will restrict ourselves to the standard drag
measuring geometry assuming zero tunneling between the layers. We
expect that allowing small tunneling will stregthen the effect; we
will pursue this possiblity in future work.

This paper is organized as follows. In Sec. II, we extend the quantum
vortex formalism to bilayers, and then we calculate the drag
resistance in the insulating and the metallic regime,
respectively. The effect of unpaired electrons on the drag resistance
is also studied. In Sec. III, we review the percolation theory of
Ref. \onlinecite{meir}, and then extend this theory to
bilayers as well, in order to calculate the drag resistance. In Sec. IV, we briefly
discuss the drag resistance behavior within the phase glass model of
Refs. \onlinecite{Phillips2002,WuPhillips}. Finally, we summarize and
discuss our results in Sec.V. Some details are provided in appendices.

\section{Drag resistance in the quantum vortex paradigm}

\subsection{The vortex description of double-layer amorphous films}\label{sec:bilayer}

Within the quantum vortex paradigm, the insulating phase has been
explained as a superfluid of vortices by the "dirty boson" model of
Ref. \onlinecite{Fisher1990}, while the metallic phase is expected to
be an uncondensed vortex liquid (see also
Ref. \onlinecite{Feigelman1993}). This picture has been pursued by
Ref. \onlinecite{vortexmetal} which argues that vortices form a Fermi
liquid for a range of magnetic field, thereby explaining the metallic
phase. At larger fields, where the insulating phase breaks down, it is claimed that
gapless bogolubov quasi particles nicknamed spinons, i.e., unpaired fermions with finite density of states at the Fermi energy, become mobile, impede
vortex motions, destroy the insulating phase, and suppress the resistance down to
normal metallic values.

We will concentrate on the case where no interlayer Josephson coupling
exists, and the vortex drag comes from the
magnetic coupling between vortices in different layers which tends to
align themselves vertically to minimize the magnetic energy. To
calculate the drag resistance in a bilayer setup, it is crucial to
derive the vortex interaction potential due to the current-current
magnetic coupling between the layers,
which is captured by the $B^2$ term in the Maxwell action. We achieve this by both a field theory
formalism and a classical calculation. The classical calculation is
relegated to Appendix \ref{vortex_interaction}. 

Let us next derive the vortex action. Treating the
superconducting film as a Cooper pair liquid, we have the following
partition function
\be
\mZ=\int\mD\rho_1\mD\rho_2\mD\theta_1\mD\theta_2\mD\A e^{-S},
\ee
where
\ben
S&=&\int_0^{\beta}\dd\tau\left\{\int \dd^2r\sum_{n=1,2}\hbar\rho_n\partial_{\tau}\theta_n+H_0+H_{int}\right\},\nonumber\\
H_0&=&\int \dd^2r\sum_{n=1,2}\frac{\rho_s}{2\hbar^2}\left(\hbar\nabla\theta_n-\frac{2e}c\A_{ext}-\frac{2e}c\A\right)^2\nonumber\\
&+&\frac1{4\pi}\int \dd^3r \vec B^2,\nonumber\\
H_{int}&=&\int \dd^2r\int \dd^2r'\frac12\sum_{n=1,2}\rho_n(r)V_i(r-r')\rho_n(r')\nonumber\\
&+&\rho_1(r)V_e(r-r')\rho_2(r'),\nonumber
\een
where $a$ is the (center-to-center) layer-separation, $\rho_n$ and $\theta_n$ are the 2d density and phase fluctuation of the $n-$th layer Cooper pair field, respectively, $A$ and $A_{ext}$ are the fluctuating and external part of the electromagnetic field, respectively. The intralayer Coulomb interaction $V_i(r)=(2e)^2/r$ (whose 2d Fourier transform would be $2\pi(2e)^2/q$), and the interlayer Coulomb interaction $V_e(r)=(2e)^2/\sqrt{r^2+a^2}$ (whose 2d Fourier transform is $2\pi(2e)^2/qe^{-qa}$). $\rho_s$ is the superfluid phase stiffness of each layer, which can be determined approximately from the Kosterlitz-Thouless temperature $T_{KT}$:
\be
T_{KT}=\frac{\pi}2\rho_s.
\ee
Next, we follow a procedure of vortex-boson duality transformation taking into account the $B^2$ term (which will be the origin of the interlayer vortex interaction), and obtain the following dual action for the vortex field $\psi_{vn}$ of the $n$-th layer and two U(1) gauge fields $\alpha_{\mu}$ and $\beta_{\mu}$ (see Appendix \ref{app:bilayer} for details):
\begin{align}\label{vortex_bilayer_action}
 S&=\sum_{\q,\o}\left\{\sum_{n=1,2}\left[-i\hbar\delta\rho_{vn}\o\phi_n
+\frac12\delta\rho_{vn}U_i\delta\rho_{vn}
\right.\right.\nonumber\\
&+\left.\frac{1}{2m_v}\left(\left(\hbar\q-e_1^*\frac{\vec{\alpha}}{c^*_1}+(-1)^ne_2^*\frac{\vec{\beta}}{c^*_2}\right)\psi_{vn}\right)^2\right]\nonumber\\
&+\delta\rho_{v1}U_e\delta\rho_{v2}+\frac1{4\pi}(\o^2-c_{*1}^2q^2)\left(\frac{\vec \alpha}{c^*_1}\right)^2\nonumber\\
&+\left.\frac1{4\pi}(\o^2-c_{*2}^2q^2)\left(\frac{\vec \beta}{c^*_2}\right)^2\right\},
\end{align}
where $\delta\rho_{vn}=\rho_{vn}-B/\Phi_0$, $\Phi_0$ is the flux
quantum, $\rho_{vn}=\psi_{vn}^{\dagger}\psi_{vn}$, $\phi_n$ is the
phase of the vortex field $\psi_{vn}$, and $m_v$ is the vortex
mass. Since there is still controversy over the theoretical value of
$m_v$, we chose to determine the vortex mass from experiments. As
discussed in Appendix \ref{duality}, for the InO film of Ref. \onlinecite{shahar2004}, we obtain $m_v\approx 19m_e$ where $m_e$ is the bare electron mass.

$\alpha_{\nu}$ and $\beta_{\nu}$ are gauge fields which mediate the
symmetric and antisymmetric part of the vortex-vortex
interaction. They are related to the Cooper pair currents $j_{n\mu}$ in the $n-$the layer by
\ben\label{dual_field_bilayer}
j_{1\mu}+j_{2\mu}&=&\frac{e_1^*}{\pi\hbar}\epsilon_{\mu\nu\eta}\partial_{\nu}\alpha_{\eta},\nonumber\\
j_{1\mu}-j_{2\mu}&=&\frac{e_2^*}{\pi\hbar}\epsilon_{\mu\nu\eta}\partial_{\nu}\beta_{\eta}.
\een

For $\nu=1,2$, the dual charges and the dual "light speeds" are
\ben\label{dual_c}
e_{\nu}^*&=&\sqrt{\pi\rho_s}\sqrt{\frac{q}{q+q_c(1-(-1)^ne^{-qa})}},\\
c^*_{\nu}&=&c\sqrt{\frac{q_c(1-(-1)^ne^{-qa})}{q+q_c(1-(-1)^ne^{-qa})}},
\een
where $q_c$ is the inverse of the 2d Pearl screening length\cite{Pearl1964}, which can be estimated from the value of $T_{KT}$:
\be
q_c=\frac{d}{2\lambda^2}=\frac{2\pi \rho_s(2e)^2}{\hbar^2c^2}=\frac{16e^2T_{KT}}{\hbar^2c^2}.
\ee
For example, the film in Ref. \onlinecite{shahar2004} has $T_{KT}$ around 0.5K. This corresponds to $q_c\approx(4$cm$)^{-1}$, and it is much smaller than the inverse of typical sample size $1/L\sim$1mm$^{-1}$.

In (\ref{vortex_bilayer_action}), we have chosen the transverse gauge
for the gauge fields $\alpha_{\mu}$ and $\beta_{\mu}$ and integrated
out $\alpha_0$ and $\beta_0$ to obtain the vortex interaction
potentials. The intralayer vortex interaction potential
\begin{align}\label{Ui}
U_i(q)
&=\frac{\Phi_0^2q_c}{2\pi}\frac{q+q_c}{q(q^2+2q_cq+q_c^2(1-e^{-2qa}))},
\end{align}
and the interlayer vortex interaction potential
\begin{align}\label{Ue}
U_e(q)&=-\frac{q_c}{q+q_c}e^{-qa}U_i.
\end{align}
When $r<1/q_c$, $U_i(r)$ gives the
familiar log interaction; for $r>1/q_c$, i.e., beyond the Pearl
screening length, $U_i(r)$ is still
logarithmic but with half of the magnitude \cite{Blatter2005}, in
contrast to the $1/r$ behavior of the single layer case (which is
Eq. (\ref{Ui}) with $a\rightarrow\infty$). The interlayer interaction
$U_e$ is purely due to the magnetic coupling, i.e., vortices in
different layers tend to align to minimize the energy cost in the
$B^2$ term. As expected, the interaction between two vortices
with the same vorticity in different layers is attractive, although
its strength is suppressed with increasing distance $a$ and decreasing
$q_c$. $U_i$ and $U_e$ can also be derived classically by solving
London equations and Maxwell's equations, which we will show in
Appendix \ref{vortex_interaction}. In addition, the form of $U_e$ is
equivalent to those derived in
Ref. \onlinecite{Sherrill1973, Sherrill1975}.

Following Ref. \onlinecite{Feigelman1993}, one can examine the strength of the interaction between vortices and transverse gauge field modes by looking at the dimensionless coupling constant
\be
\alpha_{T}\equiv\frac{e_{*1,2}^2}{m_vc_{*1,2}^2}\sim\frac{\rho_s}{m_vc^2}\cdot\frac{q}{q_c(1\pm e^{-qa})}\leq 10^{-5}
\ee
for the entire range $0\leq q\leq 1/\xi$, $\xi\sim10$nm being the coherence length. Thus, the transverse gauge field excitations can be neglected. For a comparison, the dimensionless parameter for the strength of the longitudinal interactions $U_i$ and $U_e$ is
\be
\alpha_L\equiv\frac{e_{*1,2}^2m_v}{\hbar^2 n_v}\sim\frac{\rho_sm_v}{\hbar^2 n_v}\cdot\frac{q}{q+q_c(1\pm e^{-qa})}\leq\frac{\rho_sm_v}{\hbar^2 n_v}\sim1.
\ee

With these simplification, we now rewrite the action for the bilayer system as
\be
\begin{aligned}\label{finalBilayerAction}
S&=\sum_{\q,\o}\left[-\delta\rho_{v1}i\hbar\o\phi_{1}
-\delta\rho_{v2}i\hbar\o\phi_{2}\right.\\
&+\frac12\delta\rho_{v1}U_i\delta\rho_{v1}
+\frac12\delta\rho_{v2}U_i\delta\rho_{v2}
+\delta\rho_{v1}U_e\delta\rho_{v2}\\
&+\left.\frac{1}{2m_v}\left(\hbar\q\psi_{v1}\right)^2+\frac{1}{2m_v}\left(\hbar\q\psi_{v2}\right)^2\right].
\end{aligned}
\ee
As the magnetic fields increases, $\alpha_L$ gets suppressed, and
therefore the vortex system goes from a interaction-dominated
localized phase (Cooper-pair superfluid phase, i.e., superconducting)  to a kinetic-energy-dominated
superfluid phase (Cooper-pair insulating phase), possibly through a metallic phase.
Finally, when the applied magnetic field is large enough that unpaired
electrons (``spinons`` in Ref. \onlinecite{vortexmetal}) are
delocalized, they impede vortex motion through their statistical
interaction with vortices and therefore suppress the resistance down
to values consistent with a normal state in the absence of pairing (see Ref. \onlinecite{vortexmetal}).

\subsection{Drag resistance in the vortex metal regime}\label{sec:metal}

As explained in the introduction, essentially all films undergoing a
magnetic field driven SIT also exhibit the saturation of their
resistance at the transition.  Within the vortex picture, the intervening metallic phase is interpreted
as a liquid of uncondensed vortices \cite{vortexmetal}, and the
vortices are diffusive, and have dissipative
dynamics. At intermediate fields and low temperatures, where the intermediate metallic
phase appears, the vortices are delocalized
but uncondensed. In this phase one can derive the following form of the the drag {\it
  conductance} $\sigma_D$ (which for the vortices is the equivalent through duality to the drag
resistance of charges) using either the Boltzman equation or
diagrammatic techniques, irrespective of the effective statistics of
vortices\cite{Gramila,Jauho1993,Zheng,Oreg,Hu1995,vonOppen2001,Hwang2003}:
\begin{equation}\label{2DEG_drag}
\sigma_D=\frac{\hbar^2}{8\pi^2T}\frac{\partial\sigma_1}{\partial n_1}\frac{\partial\sigma_2}{\partial n_2}\int_0^{\infty}q^3\dd q\int_0^{\infty}
\dd\omega\frac{|U|^2\im\chi_1\im\chi_2}{\sinh^2\left(\frac{\hbar\omega}{2T}\right)},
\end{equation}
where $\sigma_i$, $n_i$, and $\chi_i$ are the {\it conductance}, density, and the density response function of the vortices in the $i-$th layer. In addition,
\be
U=\frac{U_e}{(1+U_{i}\chi_1)(1+U_{i}\chi_2)-U_e^2\chi_1\chi_2}
\ee
is the screened interlayer interaction, $U_e$ is the
bare interlayer interaction, and $U_{i}$ is the
intralayer interaction, and $T$ is the temperature. $\partial\sigma_v/\partial n_v$ appears since $R_D$ is related to the
single layer rectification function, $\Gamma$, defined as $\vec j_v=\Gamma
\phi^2$, with $\phi$ being the vortex potential field. $\Gamma$ is generally
proportional to $\partial\sigma_v/\partial n_v$ (see
Ref. \onlinecite{vonOppen2001}). Combining the vortex density expression $n_i=B/\Phi_0$ and the relation between physical resistance and the vortex conductance $R=(\frac{h}{2e})^2\sigma_v$ with (\ref{2DEG_drag}), one obtains the drag resistance
\be\label{fermionic_vortex_drag}
R_D=\frac{e^2\Phi_0^2}{8\pi^4T}\frac{\partial R_1}{\partial B}\frac{\partial R_2}{\partial B}\int_0^{\infty}q^3\dd q\int_0^{\infty}
\dd\omega|U|^2\frac{\im\chi_1\im\chi_2}{\sinh^2\left(\frac{\hbar\omega}{2T}\right)}.
\ee
Remarkably, the drag resistance is
proportional to $\partial R_{1,2}/\partial B$, and thus $R_D$ peaks
when the MR attains its biggest slope. This is one of the most
important results of our analysis. Intuitively, the dependence of the
drag on $\partial \sigma_V /\partial n_V=\partial R_{1,2}/\partial B$
arises since the drag effect is the result of the nonuniformity of the
relevant particle density; how this nonuniformity affects the voltage
drop in the medium both in the primary and secondary layers is exactly
the origin of the square of the magneto-resistance slope.

The only model-dependent input is the
density response function $\chi_{1,2}$. We have computed the drag resistance using two different choices of $\chi_{1,2}$. In the remainder of this section, we follow the vortex Fermi liquid description for the
metallic phase of Ref. \onlinecite{vortexmetal} and use the fermionic response function for $\chi_{1,2}$; in Appendix \ref{appendix:hard-disc}, we treat the metallic phase as a classical hard-disk
liquid of vortices\cite{Leutheusser1982,Leutheusser1983} and use its response function accordingly for $\chi_{1,2}$. It turns out that the drag resistance results are remarkbly close for these two approaches, hence showing the robustness of our results.

If we treat vortices as fermions in this phase\cite{vortexmetal}, we use the Hubbard approximation form
for $\chi_{1,2}$ considering the short-range repulsion between
vortices and also the low density of this vortex Fermi
liquid\cite{Hwang2003, Mahan}:
\be\label{HA}
\chi(\q,\omega)=\frac{\chi_0(\q,\omega)}{1-U_i(\q)\chi_0(\q,\omega)G(\q)},
\ee
 where $G(\q)=q^2/(q^2+k_F^2)$, and $k_F$ of the vortex Fermi liquid can be easily calculated from the vortex density:
\be
k_F=\sqrt{4\pi n_v}=\sqrt{4\pi\frac{B}{\Phi_0}}.
\ee
One can define the mean free path $l$ and the transport collision time $\tau$ for vortex Fermi liquid. Their value can be estimated by combining the expression for vortex conductivity $\sigma_v=n_v\tau/m_v$ and the relation between the physical resistance and the vortex conductance $R=(\frac{h}{2e})^2\sigma_v$:
\begin{align}
\tau&=R\frac{m_v}{n_v}\left(\frac{2e}{h}\right)^2,\nonumber\\
l&=\frac{R}{\pi^2\hbar/e^2}\sqrt{\frac{4\pi}{n_v}}.
\end{align}
When $ql>1$ or $\omega \tau>1$ we approximate $\chi_0$ by the noninteracting ballistic fermion result\cite{Stern1967}:
\be
\chi_0=\nu\left(1-C_+\sqrt{|s_+|}-C_{-}\sqrt{|s_-|}\right),
\ee
where
\ben
s_+&\equiv&\left(\frac{k_F}q\right)^2-\left(\frac{m_v\omega+q^2/2}{q^2}\right)^2;\nonumber\\
s_-&\equiv&\left(\frac{k_F}q\right)^2-\left(\frac{m_v\omega-q^2/2}{q^2}\right)^2,
\een
and
\ben
C_{\pm}&=&\sgn\left(\frac{q^2}{2m_v}\pm\omega\right), \textrm{ if }s_{\pm}<0,\nonumber\\
C_{\pm}&=&\pm i,\textrm{ if }s_{\pm}>0.
\een
For $ql<1$ and $\omega\tau<1$, we use the diffusive Fermi liquid result:
\be
\chi_0=\nu\frac{Dq^2}{Dq^2-i\omega}
\ee
Plugging (\ref{HA}) into (\ref{fermionic_vortex_drag}), one can numerically compute the drag resistance. The result is given in Sec. \ref{sec:vortex_drag_result}.

Note that this result does not
crucially depend on choice of fermionic density response function above. As stated earlier, as long as vortices form an uncondensed liquid, (\ref{fermionic_vortex_drag}) remains valid. We have also
computed $R_D$ by modeling the metallic phase as a classical hard-disk
liquid of vortices\cite{Leutheusser1982,Leutheusser1983}, and putting the corresponding density response function into (\ref{fermionic_vortex_drag}). The resulting magnitude
and the behavior of $R_D$ are extremely close to the results we
obtained above within the vortex Fermi liquid frameworks (see Appendix \ref{appendix:hard-disc}). This demonstrates the universality of our results.

\subsection{Drag resistance in the insulating (vortex superfluid) regime}

According to the vortex theory, the insulating phase is a superfluid of bosonic vortices. In this regime, the vortex dynamics is presumably nondissipative. A mechanism of nondisspative supercurrent drag between bilayer bosonic superfluid systems has been studied by Ref. \onlinecite{duan1993,terentjev,Fil}. Here, we apply this approach to the superfluid of vortices in the insulating regime. In the absence of current bias, we have the following action from (\ref{finalBilayerAction}) deep in the insulating phase:
\begin{align}\label{supercurrent_drag_action}
S&=\sum_{\q,\omega}\left\{-i\dr_1\phi_1\omega+\frac{n_v}{2m_v}(-q^2\phi_1^2)\right.\nonumber\\
&-i\dr_2\phi_2\omega+\frac{n_v}{2m_v}(-q^2\phi_2^2)\nonumber\\
&+\left.\frac12U_i(\dr_1)^2+\frac12U_i(\dr_2)^2+U_e\dr_1\dr_2\right\}.
\end{align}

Switching to the canonical quantization formalism and using mean field approximation for the quartic interaction term\cite{terentjev}, the above action (\ref{supercurrent_drag_action}) corresponds to the following Hamiltonian for bilayer interacting bosons:
\begin{align}\label{supercurrent_drag_Hamiltonian}
H&=\sum_{s=\pm}\sum_{\q}\left\{\frac{q^2}{2m_v}a_s^{\dagger}(\q)a_s(\q)+\frac{n_v}2[U_i(q)+sU_e(q)]\right.\nonumber\\
&\times\left.[a_s^{\dagger}(\q)a_s^{\dagger}(-\q)+a_s(-\q)a_s(\q)]\right\},
\end{align}
where
\be
a_{\pm}(\q)=\frac1{\sqrt2}[\psi_{v1}(\q)\pm \psi_{v2}(\q)],
\ee
$\psi_{v1}$ and $\psi_{v2}$ are the bosonic vortex field operators for the first and second layer, respectively. (\ref{supercurrent_drag_Hamiltonian}) can be diagonalized using Bogoliubov transformations:
\be\label{Bogo}
a_{\pm}(\q)=u_{\pm}(\q)b_{\pm}(\q)+v_{\pm}(\q)b_{\pm}^{\dagger}(-\q),
\ee
where in the long wavelength limit
\begin{align}
u_{\pm}^2(\q)&=\frac12\left\{\frac{n_v[U_i\pm U_e]}{\o_{\pm}(q)}+1\right\},\nonumber\\
v_{\pm}^2(\q)&=\frac12\left\{\frac{n_v[U_i\pm U_e]}{\o_{\pm}(q)}-1\right\},\nonumber\\
\o_{\pm}(\q)&=\sqrt{\frac{q^2n_v}{m_v}[U_i(q)\pm U_e(q)]}.\label{spectrum2}
\end{align}
A vortex current bias $\vec{v_1}$ in layer 1 (the driving layer) is represented by a perturbation term $H_1$ in our Hamiltonian:
\be
H_1=\sum_{\q}m_v\j_1\cdot\vec{v_1}.
\ee
The drag current in the second layer can be calculated using standard perturbation theory. The new ground state to the first order in $v_1$ is given by
\be
|\O\ra=|0\ra-\sum_{n\neq0}\frac{ | n\ra\la n|H_1|0\ra}{E_n-E_0},
\ee
where $|0\ra$ is the vacuum state of $b^{\dagger}_{\pm}$, and $|n\ra$ represents all possible states obtained by acting $b^{\dagger}_{\pm}$ on $|0\ra$.
Thus, at this order,
\begin{align}
\la\j_2\ra&=\la0|\j_2|0\ra-\sum_{n\neq0}\frac{ \la 0|H_1| n\ra\la n|\j_2|0\ra}{E_n-E_0}\\
&-\sum_{n\neq0}\frac{ \la0|\j_2| n\ra\la n|H_1|0\ra}{E_n-E_0}.\nonumber
\end{align}
It is straightforward to check that the only excited states $|n\ra$ that contribute to the sum are of the form $b_+^{\dagger}(\q)b_-^{\dagger}(-\q)|0\ra$. One thus obtains
\begin{align}
\la\j_2\ra&=\frac{\vec{v_1}}{4m_v}\sum_{\q}q^2\frac{[v_+(\q)u_-(\q)-v_-(\q)u_+(\q)]^2}{\o_+(\q)+\o_-(\q)}\nonumber\\
&=\frac{\vec{v_1}}{16m_v}\sum_{\q}q^2\frac{[\o_+^2(\q)-\o_-^2(\q)]^2}{\o_+(\q)\o_-(\q)[\o_+(\q)+\o_-(\q)]^3}.\label{current_Hamiltonian}
\end{align}
Now, plugging (\ref{spectrum2}) into (\ref{current_Hamiltonian}), to the second order in interlayer interaction $U_e$ we have
\ben
\la\j_2\ra
&=&\vec{v_1}\frac{\hbar}{128a^2\Phi_0}\sqrt{\frac{q_c^3}{2\pi n_vm_v}}.\nonumber
\een
Divding this result by $\la \j_1\ra=n_v\vec{v_1}$ and recalling that the resistance is proportional to the vortex current, one is ready to obtain the drag resistance,
\be\label{jj_drag}
\frac{R_D}R=\frac{\la j_2\ra}{\la j_1\ra}=\frac{\hbar}{128a^2\Phi_0}\sqrt{\frac{q_c^3}{2\pi m_vn_v^3}}.
\ee
When spinons are mobile, they will suppress the drag resistance, as we will show in section \ref{spinon}.

\subsection{The effect of mobile spinons}\label{spinon}

The discussions in previous sections apply to the case where no mobile unpaired electrons, i.e. spinons in Ref. \onlinecite{vortexmetal}, exist in the system. However when the magnetic field is strong enough to pull apart Cooper pairs and delocalize spinons, as is signaled by the downturn of the magnetoresistance, the drag resistance is modified by the spinons. In this subsection, we analyze how mobile spinons affect our drag resistance results above.

We follow the semiclassical Drude formalism as in Ref. \onlinecite{vortexmetal} which takes into account the statistical interaction between Cooper pairs, vortices, and spinons. Vortices and spinons see each other as $\pi$-flux source, while electric current exerts Magnus force on vortices. Denoting the electric current, vortex current, and the spinon current in the $n-$th layer as $\vec J_{n}$, $\j_{v,n}$, $\j_{s,n}$, we have the following equations for the first (driving) layer (see Ref. \onlinecite{vortexmetal}):
\begin{align*}
\j_{v1}&=\sigma_v\hat{z}\times(\j_{s1}-\vec J_{1}),\\
\j_{s1}&=\sigma_s\hat{z}\times{\j_{v1}}.
\end{align*}
Similarly, denoting the vortex drag conductance without spinons as $\sigma_D$, we incorporate the drag effect in the following way in the equations of the second (passive) layer:
\begin{align*}
\j_{v2}&=\frac{\sigma_D}{\sigma_v}\j_{v1}+\sigma_v\hat{z}\times\j_{s2},\\
\j_{s2}&=\sigma_s\hat{z}\times{\j_{v2}}.
\end{align*}
This set of equations is a consequence of the absence of electric current but the presence of vortex drag effect in the second layer.
We can solve these two sets of equations, and obtain the effective vortex drag conductance:
\be
\sigma_{D}^{eff}=\frac{j_{v2}}{J_{1}}=\frac{\sigma_D}{(1+\sigma_v\sigma_s)^2}.
\ee
Since the physical resistance $R=(h/(2e))^2\sigma_v$, we have
\be
R_{D}^{eff}=\frac{R_D}{(1+R_v/R_s)^2}.
\ee
where $R_D$ is the drag resistance if spinons are localized, $R_v=(h/2e)^2\sigma_v$ is the vortex contribution to the resistance, and $R_s=\sigma_s^{-1}$ is the spinon contribution to the resistance. Thus, we see that when $R_s\ll R_v$, the drag resistance is quickly suppressed to unmeasurably small as spinon mobility increases.

\subsection{Results of the drag resistance in the vortex theory}\label{sec:vortex_drag_result}
\begin{figure}[h]
\includegraphics[scale=0.4]{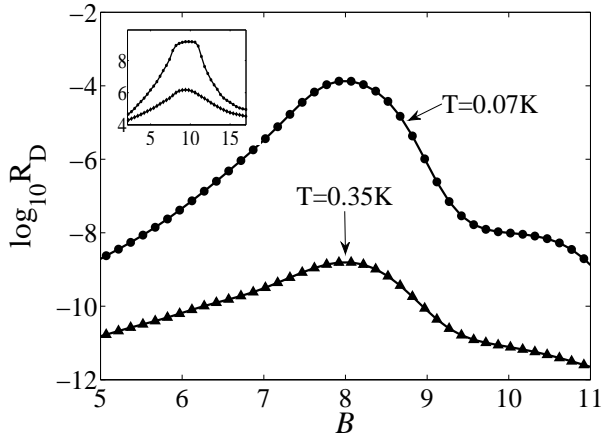}
\caption{Drag resistance $R_D$ (in Ohms) between two
identical films as in FIG.~2b of Ref. \onlinecite{shahar2004}
vs. magnetic field $B$, according to the vortex
picture\cite{vortexmetal} (log scale); . The drag resistance has been
smoothened to avoid discontinuity at the boundary between the
metallic and the insulating phase. Center-to-center layer separation
$a=25$nm, temperature $T=0.07$K and 0.35K. Insets: single layer
magnetoresistance (magnetoresistance, log scale) reproduced according to the quantum vortex theory.. The
parameters are tuned to make the magnetoresistance resemble the experimental data
in FIG. 2b of Ref. \onlinecite{shahar2004}.
$R_D$ has a peak at the steepest point
($\sim8$T) of the magnetoresistance, which is due to the fact that $R_D$ is
proportional to the square of the slope of the magnetoresistance in the small
magnetic field side of the peak. Also, $R_D$ is larger at lower
temperature, because the magnetoresistance curve is then much steeper. Carrying out the experiments at even lower
temperatures may further enhance the vortex drag effect.}\label{vortex}

\end{figure}
Collecting the above results and the value of the vortex mass $m_v$ discussed in Appendix \ref{sec:bilayer}, tuning the value of the vortex (spinon) contributions to the resistance $R_v$ ($R_s$) so that $R=R_vR_s/(R_v+R_s)$ (see Ref. \onlinecite{vortexmetal}) resembles the resistance observed in the experiment of Ref. \onlinecite{shahar2004}, and setting temperature to be 0.07K and 0.35K, we have calculated the drag resistance between two identical films with single layer resistance given by the inset of FIG.\ref{vortex}, and with center-to-center layer separation 25nm. We assume that vortices form a Fermi liquid (thus (\ref{fermionic_vortex_drag}) is applicable; however see also Appendix \ref{appendix:hard-disc}) when $B<9$T, and a bosonic superfluid (thus (\ref{jj_drag}) is used) when $B>9$T. We smoothen the drag resistance curve by convoluting it with a Gaussian function to avoid discontinuity across the phase boundary between the metallic phase and the insulating phase.

The results of vortex drag are summarized in FIG.\ref{vortex}. One can see that The drag resistance has a peak at the steepest point ($\sim8$T) of the magnetoresistance. This is due to the fact that in the vortex metal regime, the drag resistance is proportional to the square of the slope of the magnetoresistance. Also, the drag resistance is larger at lower temperature. This is because the magnetoresistance curve is much steeper as one approaches zero temperature(see (\ref{fermionic_vortex_drag})). For the film of Ref. \onlinecite{shahar2004}, the {\it sheet} drag resistance is about $10^{-1}$ m$\Omega$ at its maximum, which is measurable despite challenging. We suggest to carry out experiments to even lower temperature, which should leads to a larger drag resistance. Using a Hall-bar shape sample would also amplify the result.


\section{drag resistance in the percolation picture}

\subsection{Review of the percolation picture of the magnetoreistance}

Within the percolation picture of Ref. \onlinecite{meir}, it is argued that the non-monotonic
magnetoresistance arises from the film breaking down to superconducting and normal regions (described as localized electron glass)
\cite{meir}. As the magnetic field increases, the superconducting region shrinks,
and a percolation transition occurs. Once the normal regions
percolate, electrons must try to enter a superconducting island in
pairs, and therefore encounter a large Coulomb blockade absent in normal puddles. The magnetoresistance peak thus
reflect the competition between electron transport though narrow normal regions, and the tunneling through
superconducting islands.

This picture is captured using a
resistor network description. Each site of the network has a probability $p$ to be
normal, and $1-p$ to be superconducting; each link is assigned a
resistance from the three values $R_{NN},\,R_{SS},\,R_{SN}$, that
reflect whether the sites the link connects are
normal (N), or superconducting (S). An increase of the magnetic field is
assumed to only cause $p$ to increase. Since the normal region is described as disordered electron glass, $R_{NN}$, the resistance between two normal sites, is assumed to be of the form of hopping conduction:
\be
R_{ij}\sim R_{N0}\exp{\left(\frac2{\xi_{loc}}+\frac{|\epsilon_i|+|\epsilon_j|+|\epsilon_i+\epsilon_j|}{k_BT}\right)},
\ee
where $\xi_{loc}$ is the localization length, and $\epsilon_i$ is the
energy of the $i-$th site measured from the chemical potential (taken
from a uniform distribution $[-W/2,W/2]$), and for simplicity we allow
only nearest neighbor hopping. The resistance between two
superconducting sites, $R_{SS}$, is taken to be very small, but still
nonzero, and vanishes as $T\sim T^{\alpha} \rightarrow 0$. Most importantly, the
resistance between one normal site and a neighboring superconducting
site, $R_{SN}$, is assumed activated:
\be
R_{SN}\sim R_{SN0}\exp\left(\frac{E_c}{k_BT}\right)
\ee
to model the charging energy electrons need to pay to enter a superconducting island. 

We have reproduced the work of Ref. \onlinecite{meir} where the
parameters of this model are chosen to reproduce the
magneto-resistance curves and temperature dependence observed in the
strong-insulator InO sample \cite{shahar2004}. The total resistance vs. the
probability of normal metal (assumed to increase with increasing
magnetic field) is shown in the inset of
FIG.\ref{percolation}. Indeed, the peak of the magnetoresistance can
be explained by this theory. However, as we demonstrate now, this
theory predicts a very different behavior for the drag resistance.


\subsection{Calculation of drag resistance within the percolation picture}

\begin{figure}[h]
\includegraphics[scale=0.4]{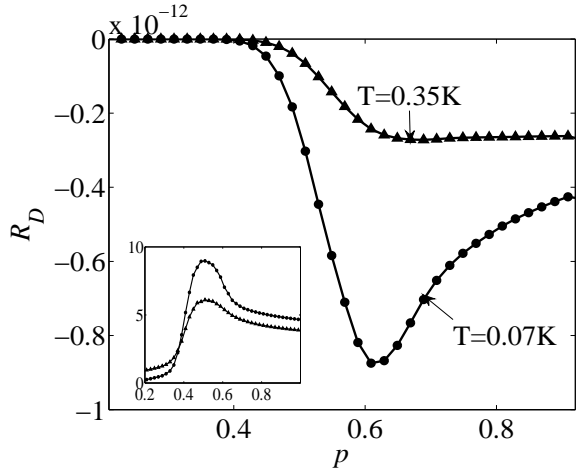}
\caption{Drag resistance $R_D$ (in Ohms) between two
identical films as in FIG.~2b of Ref. \onlinecite{shahar2004} vs. normal metal
percentage $p$ (corresponding to normal magnetic field), according to the
percolation picture\cite{meir}. Center-to-center layer separation
$a=25$nm, temperature $T=0.07K$ and $0.35K$. Insets: single layer
magnetoresistance (magnetoresistance, log scale) reproduced according to the percolation theory. The
parameters are tuned to make the magnetoresistance resemble the experimental data
in FIG. 2b of Ref. \onlinecite{shahar2004}.
The sign of the voltage drop of the passive layer is opposite
to that of the driving layer, and the maximum magnitude value of $R_D$ is
much smaller, $\sim10^{-12}\Omega$.}\label{percolation}
\end{figure}

To calculate $R_D$, we first follow Ref. \onlinecite{meir} and tune the parameters to make
the single layer resistance resemble the experimental data in
FIG. 2(b) of Ref. \onlinecite{shahar2004}: $\xi_{loc}=0.1$, $W=0.4$K, $E_c=0.6$K, $R_{SN0}\sim10^{6}\Omega$, and $R_{N0}\sim10^{-5}\Omega$. Next, we place one such
network (active layer) on top of another one (passive layer). Each
link is treated as a subsystem, which might induce a drag voltage (an
emf) $\varepsilon=I R_D$ in the link under it in the passive
layer. When a link is between two normal (or superconducting) sites, it is
treated as a disorder localized electron glass (or superconductor). In Appendix \ref{Coulomb_drag}, we find $R_D$ between
two localized electron glass separated by vacuum is:
\be\label{PercolationDrag}
R_D\approx\frac1{96\pi^2}\frac{R_1R_2}{\hbar/e^2}\frac{T^2}{(e^2nad)^2}\ln\frac1{2x_0}.
\ee
Here, $n\approx5\times10^{20}$cm$^{-3}$ is the typical carrier density
of InO\cite{Kapitulnik2005}, $d=20$nm is the film thickness, $a=25$nm
is the center-to-center layer separation, $R_{1,2}$ are the
resistances of the two normal-normal(NN) links, $x_0=a/(2\pi e^2\nu
d\xi^2)$ where $\nu$ is the density of states and $\xi\approx1$nm is
the localization length. The value of the localization length $\xi$ is
estimated by following Ref. \onlinecite{meir} to take
$\xi\sim0.1\times$ plaquette size (reflecting the fact that it is a
disordered insulator), and we estimate the plaquette size as the
superconducting coherence length $\sim10$nm. Although this estimation
of localization length is crude, the drag resistance $R_D$ has only
logarithmic dependence on it in (\ref{PercolationDrag}). Setting
$T=0.07$K, and $R_1=R_2=10^5\Omega$, we can estimate
$R_D\sim10^{-12}\Omega$.

On the other hand, we will show in Appendix \ref{SS_drag} that a
genuine (i.e., without mobile vortices) superconductor has no drag
effect at all in a resistor network, either when it is aligned with
another superconductor link or a normal link. Thus, drag effects
associated with a superconducting link can only come from
vortices. However, The small resistance for the superconducting islands in this theory implies that
vortices in the superconducting islands, if any, have very low
mobility. If two superconducting links are vertically aligned, we can
estimate the drag resistance due to mobile vortices using our vortex
drag result (\ref{fermionic_vortex_drag}): roughly $R_D\propto R^2$,
for $R\sim 10^9\Omega$ we obtained $R_D\sim10^{-4}\Omega$, therefore
for $R\sim 1\Omega$ we have $R_D\sim10^{-20}\Omega$, which is
negligible compared to the Coulomb drag resistance between two NN
links $\sim 10^{-12}\Omega$. Finally, Ref. \onlinecite{Narayan2003}
has shown that a current off the plane where vortices reside does not
exert any force on vortices. By Newton's third law or equivalently the
Kubo formula for the drag conductance, this also implies that moving
vortices does not exert any DC emf in another layer. Therefore, there
is no drag effect when a NN link is aligned with a SS
link. Consequently, the Coulomb drag between two vertically aligned NN
links (Eqn. (\ref{PercolationDrag})) dominates the drag effect.

Thus, we solve the Kirchoff's equations for the two
layers, and obtain the voltage drop and thereby the drag resistance. The
results are shown in FIG. \ref{percolation}, with $T=0.07$K and $0.35$K, film-thickness 20nm, and the center-to-center interlayer distance 25nm. We observe that the sign of the voltage drop of the passive layer is opposite to that of the driving layer (not shown in the Figure), as expected and explained in the introduction, and the maximum magnitude of the drag resistance is around $10^{-12}\Omega$, indeed much smaller than that in the vortex paradigm.


\section{Discussion on the drag resistance in the phase glass theory}

A third theory, namely the phase glass
theory\cite{Phillips2002,WuPhillips}, focuses on the nature of the
metallic phase intervening the superconducting and insulating
state. In this theory, the system is described as interacting bosons
(Cooper pairs), but it is argued that the glassy phase is in fact a
Bose metal, due to the coupling to the glassy landscape.

Specifically, Ref. \onlinecite{Phillips2002} has studied the quantum rotor model
\be
H=-E_c\sum_i\left(\frac{\partial}{\partial\theta_i}\right)^2-\sum_{\la i,j\ra}J_{ij}\cos(\theta_i-\theta_j),
\ee
where the Josephson coupling $J_{ij}$ obeys a Gaussian distribution
with nonzero mean. This model is appears to exhibit three phases: superconducting
phase, phase glass phase, and a Mott insulator
phase. Ref. \onlinecite{Phillips2002} has employed replica trick to
obtain the Landau theory of the the phase glass phase near the
glass-superconductor-transition critical point, and has calculated the
conductance in this regime. It was found that in this regime the DC
conductance is actually finite at zero temperature. For completeness,
we note that Ref. \onlinecite{Ikeda2007} argued against these results
and obtained infinite conductance instead.

This analysis has recently been extended to include the external
perpendicular magnetic field\cite{WuPhillips}, which is more relevant
to the experiments on the magnetic field tuned transition. However,
Ref.\onlinecite{WuPhillips} has only studied the regime of small
magnetic field where one just enters the resistive glassy phase and
left out issues such as the peak in the magnetoresistance. Therefore,
we leave a complete analysis to future work and simply observe that
according to this theory, the resistive state is a glassy phase where
phase variables $\theta_i$'s of the bosons are ordered locally. In
other words, there are no mobile vortices moving around. Consequently,
the current coupling as we considered in the vortex drag should is
absent, and the Coulomb interaction should dominate the drag
effect. Therefore, we expect that the sign of the drag voltage is
opposite to the voltage drop of the driving layer, as we discussed in
the introduction to be a general feature of the Coulomb drag, and the
magnitude of the drag resistance should be small. This is in part
because for a bosonic system, the phase space available for
excitations is much smaller than fermionic systems due to the absence
of a Fermi surface.

\section{summary and discussion}

One of the most exciting possiblities is that the SIT
in amorphous thin films realizes the vortex condensation scenario
\cite{FisherLee,Fisher1990,vortexmetal}. The amorphous-films Giaver transformer experiment \cite{drag}, would
be able to measure a distinct signature of mobile vortices, which is a
drag resistance opposite in its direction to that of coulomb drag.  
Therefore such a measurement would able to disclose whether the vortex
paradigm is suitable for explaining the complex
phase diagram of amorphous films in a normal manetic field, or whether
the percolation paradigm is indeed more appropriate. We provide
a detailed computation of the drag resistance according to the vortex theories of
Ref. \onlinecite{Fisher1990,vortexmetal} and the percolation theory of
Ref. \onlinecite{meir}. The drag resistance implied by the phase glass
model\cite{Phillips2002,WuPhillips} is also briefly discussed. We find
that vortex picture predicts a drag resistance orders of magnitude
stronger than non-vortex pictures. In addition, the drag resistance
and the single layer resistance have the same sign according to the
vortex picture, but the opposite sign for non-vortex
pictures. Therefore, drag resistance measurement are indeed able to
distinguish different theoretical paradigms qualitatively.

We considered specifically a bilayer device which will contain two identical films as in Ref. \onlinecite{shahar2004} with $25$nm
layer separation and at $0.07$K. A calculation within the vortex
paradigm yields a drag resistance
$R_D\sim10^{-4}\Omega$ at its maximum value. This drag arises solely from the
attractive interaction of the demagnetizing currents of vortices. The
value we find is probably near the
limit of measurability;  we suggest, however, to carry out experiments at
even lower temperature, in which case the single layer
magnetoresistance is even steeper, and the drag resistance should be
larger. Within the percolation picture of Ref. \onlinecite{meir}, the dominating drag effect is
the drag between two vertically aligned normal regions in the different
layers. For two identical films as in Ref. \onlinecite{shahar2004}
with $25$nm layer separation at $0.07$K, we find the drag resistance
$R_D\sim10^{-12}\Omega$ at its maximum value, which is indeed orders
of magnitude smaller than the drag resistance predicted by the vortex
picture. Also, we find the sign of the drag resistance is the opposite
of that of the single layer resistance, as expected.

The answer we find should not depend crucially on the details of the
microscopic picture which we use. If vortices are not responsible for
the inhibitive resistance which the films display, then drag effects
will appear primarlily due to Coulomb repulsion of single electrons. 
This drag effect will be low because of the relatively high electronic
density in the films. On the other hands, if vortices are responsible
for the large resistance in the intermediate magnetic fields leading
to the insulating phase, then they will produce a drag opposite in its
direction to the Coulomb drag. To carry out the vortex drag
calculation in the metallic phase intervening between the
superconducting and insulating phase we used the picture of Ref. \onlinecite{vortexmetal},
which treats the vortices as fermionic diffusive particles. This picture is
justified due to the strong long-ranged interactions within the vortex
liquid, which render the question of statistics secondary,
intuitively, since vortices rarely encircle each other. Nevertheless,
to demonstrate the universality of our results, we also carried out
the drag calculation in the metalic phase assuming that the vortices
are hard core disks, and obtained essentially the same answer
(c.f. App. \ref{appendix:hard-disc}). 

Indeed our strongest results are obtained in the intermediate-field
metallic phase. The controversy surrounding this phase requires some
special attention. First, we note that all experiments of thin
amorphous films exhibit a saturation of the resistance at temperature
below about $100$mK at intermediate resistances. This is clearly seen
in, e.g, the resistance vs. field traces which overlap at subsequent temperature
sweeps as in Fig.2b of Ref. \onlinecite{shahar2004}. Second, there are reasons to
believe that this saturation is not the result of failure to cool
electrons. Resistances that are
too low or too high continue to change as the temperature is
lowered. But the two heating mechanisms most likely are current
heating, with power $\sim I^2R$, and therefore affecting the highest
temperatures, and ambient RF heating, which would have a
voltage-biased power $\sim V^2/R$, and therefore most effective in the
lowest resistances. Neither mechanism explains resistance saturation
at intermediate temperatures. Furthermore, experiments on Tantalum
films show distinct signatures in the metallic regime
which disappear in the insulating and superconducting regimes, and
also distinguish it from the thermally-destroyed superconducting phase\cite{Yoon2005}. 
Third, even if the metallic behavior of
the films is a finite temperature phenomena, within the vortex
paradigm, the resistance still arises due to vortex
motion. Therefore the drag calculated within this paradigm using a
diffusive vortex model should still be adequate, and our results do
not depend crucially on the existence of a zero-temperature
intervening metallic state.  

The signatures we expect to find in the proposed magnetic and Coulomb drag measurements are
not large. Incorporating interlayer electron and Josephson tunneling
will increase both the vortex-drag effect and the competing Coloumb
drag effects. As we point out here, the drag signature of vortex
motion, or single electrons or
Cooper-pairs motion will have opposite signs. Quite possibly, allowing
interlayer tunneling will render both drag effects measurable. Indeed,
such a setup will be a deviation from standard drag measurements where
charge transfer between layers is forbidden. Nevertheless, a careful
choice of tunneling strength and sample geometry will make such
experiments plausible and useful. We intend to analyze the vortex and
Coloumb drag in the presence of interlayer tunneling in future work.

\acknowledgments
It is a pleasure to thank Yonatan Dubi, Jim Eisenstein, Alexander
Finkel'stein, Alex Kamenev, Yen-Hsiang Lin, Yigal Meir, Yuval Oreg, Philip Phillips, Ady Stern, Jiansheng Wu, and Ke Xu
for stimulating discussions. This work was supported by the Research
Corporation's Cottrell award (G.R.), and by NSF through grant number DMR-0239450 (J.Y.).

\appendix

\section{The determination of the vortex mass}\label{duality}

In this appendix, we demonstrate in detail the derivation of the
vortex-boson duality for a single layer and discuss the value of the
vortex mass. Our starting point is the following partition function
for Cooper pairs:
\be
\mathcal Z = \int \mD\rho \mD\theta\mD\A e^{-S},
\ee
where the action $S$ is
\ben\label{1}
S&=&\int_0^{\beta}\dd\tau\left\{\int \dd^2r(\hbar\rho\partial_{\tau}\theta+H_0+H_{int})\right\},\nonumber\\
H_0&=&\int \dd^2r\frac{\rho_s}{2\hbar^2}\left(\hbar\nabla\theta-\frac{2e}c\A_{ext}-\frac{2e}c\A\right)^2\nonumber\\
&+&\frac1{4\pi}\int \dd^3r \vec B^2,\nonumber\\
H_{int}&=&\int \dd^2r\int \dd^2r'\frac12\rho(r) V(r-r')\rho(r').
\een
Here, $\rho$ and $\theta$ are the density and phase fluctuation of the Cooper pair field, respectively, $\vec A$ is the fluctuating electromagnetic field, and $\vec A_{ext}$ is the applied external electromangetic field, typically a perpendicular magnetic field. $V(r)=(2e)^2/r$ (whose 2d Fourier transform would be $2\pi(2e)^2/k$) is the Coulomb interaction between Cooper pairs. $\rho_s$ is the bare stiffness for phase fluctuations. The value of $\rho_s$ can be determined approximately by the zero-field Kosterlitz-Thouless temperature $T_{KT}$:
\be\label{KT}
T_{KT}=\frac{\pi}2\rho_s.
\ee
The 2d number current of Cooper pairs is
\be\label{j}
\j=\frac{\rho_s}{\hbar^2}\left(\hbar\nabla\theta-\frac{2e}c\A_{ext}-\frac{2e}c\A\right).
\ee
One can introduce the dynamical field $\j$ by Hubbard-Stratonavich transformation (or Villain transformation in the lattice version of this derivation) and transform $\mZ$ to be
\be
\mZ=\int\mD\rho\mD\theta\mD\j\mD\A e^{-S},
\ee
where
\ben
S&=&\sum_{\o,\q}\left\{-i\hbar \o\rho\theta+\frac12\rho V\rho+\frac{\hbar^2}{2\rho_s}\j^2\right.\nonumber\\&+&i\j\cdot\left(\hbar(\nabla\theta)_q-\frac{2e}c\A_{ext}-\frac{2e}c\A(\q,z=0)\right)\nonumber\\
&+&\left.\int\frac{dk_z}{2\pi}\frac{q^2+k_z^2}{4\pi}\A^2(\q,k_z)\right\}.
 \een
Here, $i$ is the imaginary number unit, $\q$ is the in-plane 2d wave vector, while $k_z$ is the 3rd wave vector component perpendicular to the plane, and subscripts $\q$ mean Fourier transformed variables. Next we split the $\theta$ field into a smooth part $\theta_s$ and a vortex part $\theta_v$: $\theta=\theta_s+\theta_v$. Afterwards one can integrate out $\theta_s$ to obtain the continuity constraint:
\be
\mZ=\int\mD\rho\mD\j\mD\theta_v\mD\A\delta(\partial_t\rho+\nabla\cdot\j) e^{-S},
\ee
where
\begin{align*}
S&=\sum_{\o,\q}\left\{-i\hbar \o\rho\theta_v+\frac12\rho V\rho+\frac{\hbar^2}{2\rho_s}\j^2\right.\nonumber\\&+i\j\cdot\left(\hbar(\nabla\theta_v)_q-\frac{2e}c\A_{ext}-\frac{2e}c\A(\q,z=0)\right)\nonumber\\
&+\left.\int\frac{dk_z}{2\pi}\frac{q^2+k_z^2}{4\pi}\A^2(\q,k_z)\right\}.
\end{align*}
Furthermore, noting that $\A(\q,z=0)=\int\frac{dk_z}{2\pi}\A(\q,k_z)$, one can integrate out $\A$ in its transverse gauge, and the action $S$ now reads
\ben
S&=&\sum_{\o,\q}\left\{-i\hbar \o\rho\theta_v+\frac12\rho V\rho+i\j\cdot\left(\hbar(\nabla\theta_v)_q-\frac{2e}c\A_{ext}\right)\right.\nonumber\\
&+&\left.
\frac{\hbar^2}{2\rho_s}\left(1+\frac{q_c}q\right)\j^2\right\},
\een
where $q_c$ is the inverse of the 2d Pearl screening length\cite{Pearl1964}, and typically it is much smaller than $1/L$, where $L$ is the sample size.

The continuity constraint is solved by defining a new gauge field $a_{\mu}=(a_0,\a)$ such that
\be\label{dual_gauge_field2}
j_{\mu}=\frac1{\eta}\epsilon_{\mu\nu\eta}\partial_{\nu}a_{\eta},
\ee
where $j_{\mu}=(c^*\rho,\j)$ and $\partial_{\mu}=(\frac1{c^*}\partial_{\tau},\nabla)$, and the value of constant $\eta$ and the "speed of light" $c^*$ are to be determined.
Writing in components, (\ref{dual_gauge_field2}) is
\be\label{dual_eb}
\vec e=\eta\j\times\hat z,\qquad b=\eta c^*\rho,
\ee
where $\vec e$ and $b$ are the dual "electric field" and "magnetic field" associated with $\alpha$, respectively. To fix $\eta$ and $c^*$, we require
\be
\frac1{4\pi}\vec e^2=\frac{\hbar^2}{2\rho_s}\left(1+\frac{q_c}q\right)\j^2,\qquad \frac1{4\pi}b^2=\frac12\rho V\rho,
\ee
thus
\be
\eta\equiv\sqrt{\frac{2\pi \hbar^2}{\rho_s}\frac{q+q_c}q},\qquad
c^*=\sqrt{\frac{2\pi(2e)^2\rho_s}{(q+q_c)\hbar^2}}.
\ee
Using (\ref{dual_gauge_field2}), we express the partition function $\mZ$ as
\be
\mZ=\int\mD\a\mD a_0\mD \theta_v e^{-S},
\ee
where
\ben
S&=&\sum_{\o,\q}\left\{\frac1{\eta}\epsilon_{\mu\nu\eta}q_{\nu}a_{\eta}\left(\hbar(\partial_{\mu}\theta_v)_q-\frac{2e}cA^{ext}_{\mu}\right)\right.\nonumber\\
&+&\left.\frac1{4\pi}(\o^2-c_*^2q^2)\left(\frac{\a}c_*\right)^2+\frac{q^2}{4\pi}a_0^2
\right\}.
\een
Integrating by parts, and noting the definition of the vortex current density
\be\label{vortex_current}
j^v_{\mu}=\frac1{2\pi}\epsilon_{\mu\nu\eta}\partial_{\nu}\partial_{\eta}\theta^v,
\ee
we obtain
\ben
S&=&\sum_{\o,\q}\left\{-e^*ia_0\left(\rho_v-\frac{B_{ext}}{\Phi_0}\right)+ie^*\j^v\cdot\frac{\a}{c^*}\right.\nonumber\\
&+&\left.\frac1{4\pi}(\o^2-c_*^2q^2)\left(\frac{\a}c_*\right)^2+\frac{q^2}{4\pi}a_0^2,
\right\}.
\een
where $\Phi_0=hc/(2e)$, and the "dual charge" of vortices is
\be\label{dual_charge2}
e^*=\frac{2\pi\hbar}{\eta}=\sqrt{2\pi\rho_s}\sqrt{\frac{q}{q+q_c}}.
\ee
In the above, we have assumed that the only external electromagnetic field is a perpendicular magnetic field $B_{ext}$.

The magnitude of the Magnus force, which now appears as the electric force, can be easily verified:
\ben
F&=&e^*\times|\vec e|=\frac{2\pi\hbar}{\eta}\times\eta j=hj,
\een
as expected.

Introducing a vortex field $\psi_v$ and making the action explicitly gauge-invariant, we write the action as
\begin{align}\label{vortex_single_layer_action}
S&=\sum_{\q,\o}\left\{\delta\rho_v(-\hbar i\omega\phi-ie^*a_0)+\frac{1}{2m_v}\left[\left(\hbar\q-e^*\frac{\a}{c_*}\right)\psi_v\right]^2\right.
\nonumber\\&\qquad+\left.\frac1{4\pi}(\o^2-c_*^2q^2)\left(\frac{\a}{c_*}\right)^2+\frac{q^2}{4\pi}a_0^2\right\},
\end{align}
where $\delta\rho_v=\rho_v-\frac{B_{ext}}{\Phi_0}$, and we have introduced the vortex mass $m_v$. Integrating out $a_0$, one obtains
\ben\label{action_single}
S&=&\sum_{\q,\o}\left\{-\delta\rho_v\hbar i\omega\phi+\frac12\delta\rho_v U\delta\rho_v\right.\nonumber\\
&+&\frac{1}{2m_v}\left[(\hbar\q-e^*\frac{\a}{c_*})\psi_v\right]^2
\nonumber\\&+&\left.\frac1{4\pi}(\o^2-c_*^2q^2)\left(\frac{\a}{c_*}\right)^2\right\},
\een
where
\be
U(q)=\frac{\Phi_0^2q_c}{2\pi}\frac1{q(q+q_c)}
\ee
is the well-known Pearl interaction potential\cite{Pearl1964}.

In the insulating phase, i.e., the vortex condensed phase with vortex superfluid stiffness $\rho_{vs}$, we have
\ben\label{S1}
S&=&\sum_{\q,\o}\left\{-\delta\rho_v\hbar i\omega\phi+\frac{\rho_{vs}}{2\hbar^2}\left(i\hbar\q\phi-e^*\frac{\a}{c_*}\right)^2
\right.\nonumber\\&+&\left.\frac12\delta\rho_v U\delta\rho_v+\frac1{4\pi}(\o^2-c_*^2q^2)\left(\frac{\a}{c_*}\right)^2\right\}.
\een
Due to the Higgs mechanism in this "symmetry broken phase", the gap of the two modes in the vortex superfluid phase coincide to be
\be
E_{gap}=\sqrt{2\pi \rho_{vs}e_*^2}\approx2\pi\sqrt{\rho_{vs}\rho_s}
\ee
for $q_c\ll L^{-1}$. Roughly speaking the two modes correspond to a
density fluctuation of the vortices, or of the underlying Cooper-pairs
Deep in the insulating phase, i.e., near the peak of the magnetoresistance, the vortex stiffness is simply
\be
\rho_{vs}=\hbar^2\frac{n_v}{m_v},
\ee
where the vortex density $n_v\equiv B/\Phi_0$. Therefore, in this regime we have
\be\label{gap}
E_{gap}=2\pi\hbar\sqrt{\frac{n_v}{m_v}\rho_s}.
\ee
Since the gauge field $a_{\mu}$ is actually the fluctuation of Cooper
pairs, we conjecture that its gap $E_{gap}$ can be identified with the
activation gap observed in the experiments of Ref. \onlinecite{shahar2004,Kapitulnik2005} near the insulating peak. Ref.\onlinecite{shahar2004,Kapitulnik2005} have also found that with increasing disorder strength, the ratio $E_{gap}/T_{KT}$ is enhanced. This is natural from our expression (\ref{gap}): dividing (\ref{gap}) by (\ref{KT}), we have
\be
\frac{E_{gap}}{T_{KT}}=4\hbar\sqrt{\frac{n_v}{m_v}\frac{1}{\rho_s}};
\ee
increasing disorder makes vortices more mobile and
thereby suppresses the vortex mass $m_v$ \cite{Wallin1994}; it also suppresses the superfluid stiffness $\rho_s$. Therefore, $E_{gap}/T_{KT}$ is larger for more disordered sample.

Since there is still controversy over its theoretical value, we chose to use the experimental value of $E_{gap}$ as an input to deduce the vortex mass from (\ref{gap}). Combining (\ref{KT}), we can express the vortex mass $m_v$ as a function of observable quantities:
\be\label{m_v}
m_v=\frac{8\pi n_vT_{KT}}{E_{gap}^2}.
\ee
Again, the vortex density $n_v=B/\Phi_0$. For the InO film of Ref. \onlinecite{shahar2004}, $T_{KT}\approx0.5$K, and $E_{gap}\approx1.6$K at $B=9$T. Plugging these into (\ref{m_v}), we obtain $m_v\approx 19m_e$ where $m_e$ is the bare electron mass. For comparison, this value is not far from that of the so-called core mass of dirty superconductors\cite{KL1975,Blatter1991,DuanLeggett1992,Sonin1998} $m\sim (k_Fd)m_e\sim49m_e$ if we use carrier density $\sim 5\times10^{20}$cm$^{-3}$ and $d\sim20$nm (see Ref. \onlinecite{shahar2004,Kapitulnik2005}).

\section{The field theory derivation of the vortex interaction
  potentials in bilayers}\label{app:bilayer}

For identical bilayer superconducting thin films separated by a (center-to-center) distance $a$, we have the following partition function for Cooper pairs:
\be
\mZ=\int\mD\rho_1\mD\rho_2\mD\theta_1\mD\theta_2\mD\A e^{-S},
\ee
where
\ben
S&=&\int_0^{\beta}\dd\tau\left\{\int \dd^2r\sum_{n=1,2}\hbar\rho_n\partial_{\tau}\theta_n+H_0+H_{int}\right\},\nonumber\\
H_0&=&\int \dd^2r\sum_{n=1,2}\frac{\rho_s}{2\hbar^2}\left(\hbar\nabla\theta_n-\frac{2e}c\A_{ext}-\frac{2e}c\A\right)^2\nonumber\\
&+&\frac1{4\pi}\int \dd^3r \vec B^2,\nonumber\\
H_{int}&=&\int \dd^2r\int \dd^2r'\frac12\sum_{n=1,2}\rho_n(r)V_i(r-r')\rho_n(r')\nonumber\\
&+&\rho_1(r)V_e(r-r')\rho_2(r'),\nonumber
\een
where $\rho_n$ and $\theta_n$ are the density and phase fluctuation of the $n-$th layer Cooper pair field, respectively, $A$ and $A_{ext}$ are the fluctuating and external part of the electromagnetic field, respectively. The intralayer Coulomb interaction $V_i(r)=(2e)^2/r$ (whose 2d Fourier transform would be $2\pi(2e)^2/q$), and the interlayer Coulomb interaction $V_e(r)=(2e)^2/\sqrt{r^2+a^2}$ (whose 2d Fourier transform is $2\pi(2e)^2/qe^{-qa}$). $\rho_s$ is the superfluid phase stiffness of each layer. 

Similar to the single layer case in Appendix \ref{duality}, we can again introduce Hubbard-Stratonavich fields $\j_{1,2}$, split $\theta$'s into smooth parts $\theta_s$ and vortex parts $\theta_v$, integrate out $\theta_s$ and $\A$, and obtain
\ben
\mZ&=&\int \mD\rho_1\mD\rho_2\mD\theta^v_1\mD\theta^v_2\mD\j_1\mD\j_2\nonumber\\
&\times&\delta(\partial_t\rho_1+\nabla\cdot\j_1)\delta(\partial_t\rho_2+\nabla\cdot\j_2)e^{-S}
\een
where
\ben
S&=&\sum_{\o,\q}\left\{-i\hbar\o\rho_1\theta^v_1+i\j_1\cdot\left(\hbar(\nabla\theta^v_1)_q-\frac{2e}c\A_{ext}\right)\right.\nonumber\\
&-&i\hbar\o\rho_2\theta^v_2+i\j_2\cdot\left(\hbar(\nabla\theta^v_2)_q-\frac{2e}c\A_{ext}\right)\nonumber\\
&+&\frac12\rho_1 V_i\rho_1+\frac12\rho_2 V_i\rho_2+\rho_1 V_e\rho_2\nonumber\\
&+&
\frac{\hbar^2}{2\rho_s}\left(1+\frac{q_c}q\right)\j_1^2+\frac{\hbar^2}{2\rho_s}\left(1+\frac{q_c}q\right)\j_2^2\nonumber\\
&+&\left.\frac{\hbar^2}{\rho_s}\frac{q_c}qe^{-qa}\j_1\cdot\j_2\right\}.
\een
The difference from the single layer case is that now the continuity constraint is solved by introducing two new gauge fields $\alpha_{\mu}=(\alpha_0,\vec\alpha)$ and $\beta_{\mu}=(\beta_0,\vec\beta)$  such that
\ben\label{alphabeta}
j_{1\mu}+j_{2\mu}&=&\frac1{\eta_1}\epsilon_{\mu\nu\eta}\partial_{\nu}\alpha_{\eta},\nonumber\\
j_{1\mu}-j_{2\mu}&=&\frac1{\eta_2}\epsilon_{\mu\nu\eta}\partial_{\nu}\beta_{\eta};\nonumber
\een
Denoting the electric field and the magnetic field associated with $\alpha_{\mu}(\beta_{\mu})$ are $\vec e_1$ and $b_1$ ($\vec e_2$ and $b_2$), respectively, we have
\ben
\vec e_1&=&\eta_1(\j_1+\j_2)\times\hat z,\qquad b_1=\eta_1 c_{*1}(\rho_1+\rho_2)\nonumber\\
\vec e_2&=&\eta_2(\j_1-\j_2)\times\hat z,\qquad b_2=\eta_2 c_{*2}(\rho_1-\rho_2).
\een
To fix $\eta_{1,2}$ and the "speeds of light" $c_{*1,2}$, we require
\be
\begin{aligned}
&\frac1{4\pi}(\vec e_1^2+\vec e_2^2)=\frac{\hbar^2}{2\rho_s}\left(1+\frac{q_c}q\right)(\j_1^2+\j_2^2)
+\frac{\hbar^2}{\rho_s}\frac{q_c}qe^{-qa}\j_1\cdot\j_2;\nonumber\\
&\frac1{4\pi}(b_1^2+b_2^2)=\frac12\rho_1 V_i\rho_1+\frac12\rho_2 V_i\rho_2+\rho_1 V_e\rho_2,\nonumber
\end{aligned}
\ee
thus for $n=1,2$,
\ben
\eta_n&=&\sqrt{\frac{\pi \hbar^2}{\rho_s}\left(1+\frac{q_c}q\left(1-(-1)^ne^{-qa}\right)\right)},\\
c_{*n}&=&c\sqrt{\frac{q_c(1-(-1)^ne^{-qa})}{q+q_c(1-(-1)^ne^{-qa})}}.
\een
Using (\ref{alphabeta}) and (\ref{vortex_current}), we can again integrate by parts and express the partition function $\mZ$ as
\be
\mZ=\int\mD\alpha\mD \beta\mD \theta_{v1}\mD\theta_{v2} e^{-S},
\ee
where
\ben
S&=&\sum_{\o,\q}i\left\{-(e_1^*\alpha_0+e_2^*\beta_0)\left(\rho_{v1}-\frac{B_{ext}}{\Phi_0}\right)\right.\nonumber\\
&-&i(e_1^*\alpha_0-e_2^*\beta_0)\left(\rho_{v2}-\frac{B_{ext}}{\Phi_0}\right)\nonumber\\
&+&i\j_{v1}\cdot(e_1^*\frac{\vec\alpha}{c_{*1}}+e_2^*\frac{\vec\beta}{c_{*2}})+i\j_{v2}\cdot(e_1^*\frac{\vec\alpha}{c_{*1}}-e_2^*\frac{\vec\beta}{c_{*2}})\nonumber\\
&+&\frac1{4\pi}(\o^2-c_{*1}^2q^2)\left(\frac{\vec \alpha}{c_{*1}}\right)^2+\frac{q^2}{4\pi}\alpha_0^2\nonumber\\
&+&\left.\frac1{4\pi}(\o^2-c_{*2}^2q^2)\left(\frac{\vec\beta}{c_{*2}}\right)^2+\frac{q^2}{4\pi}\beta_0^2
\right\},
\een
and for $n=1,2$, the dual "charges" of the vortices are
\ben
e_n^*&=&\frac{\pi\hbar}{\eta_n}=\sqrt{\pi\rho_s}\sqrt{\frac{q}{q+q_c(1-(-1)^ne^{-qa})}},\nonumber\\
\een

When a (number) current bias $\j_1$ is applied in layer 1, the force on a vortex in this layer is
\begin{align*}
F&=e_1^*\times|\vec e_1|+e_2^*\times|\vec e_2|=e_1^*\eta_1|\j_1|+e_2^*\eta_2|\j_1|\\
&=h|\j_1|,
\end{align*}
and the force on a vortex in the other layer is
\begin{align*}
F&=e_1^*\times|\vec e_1|-e_2^*\times|\vec e_2|=e_1^*\eta_1|\j_1|-e_2^*\eta_2|\j_1|\\
&=0,
\end{align*}
as expected.

Again, introducing vortex fields $\psi_{v1}$ and $\psi_{v2}$ for each layer and making the action explicitly gauge-invariant, we can write the action as in \begin{align}\label{S_dual_bilayer}
S&=\sum_{\q,\o}\left\{\sum_{n=1,2}\left[\frac{\left(\left(\hbar\q-e_1^*\frac{\vec{\alpha}}{c^*_{1}}+(-1)^ne_2^*\frac{\vec{\beta}}{c^*_{2}}\right)\psi_{vn}\right)^2}{2m_v}\right.\right.\nonumber\\
&+\left.\delta\rho_{vn}\left(-i\hbar\o\phi_n-ie_1^*\alpha_0+(-1)^nie_2^*\beta_0\right)\right]\nonumber\\
&+\frac1{4\pi}(\o^2-c_{*1}^2q^2)\left(\frac{\vec \alpha}{c^*_{1}}\right)^2+\frac1{4\pi}(\o^2-c_{*2}^2q^2)\left(\frac{\vec \beta}{c^*_{2}}\right)^2\nonumber\\
&+\left.\frac{q^2}{4\pi}\alpha_0^2+\frac{q^2}{4\pi}\beta_0^2\right\}.
\end{align}
Integrating out $\alpha_0$ and $\beta_0$, one obtains the intralayer vortex interaction potential
\begin{align}
U_i(q)
&=\frac{\Phi_0^2q_c}{2\pi}\frac{q+q_c}{q(q^2+2q_cq+q_c^2(1-e^{-2qa}))},
\end{align}
and interlayer vortex interaction potential
\begin{align}
U_e(q)
&=-\frac{q_c}{q+q_c}e^{-qa}U_i.
\end{align}
Which concludes the field-theory derivation of the interaction potential. 

\section{Classical derivation of the vortex interaction potential}\label{vortex_interaction}

In this appendix, we present an alternative way of deriving the vortex
interaction potential between two vortices in a single superconducting
thin film and in bilayer thin films.

First, consider the current and electromagnetic field configuration of a single vortex at $r=0$ in a single superconducting thin film with thickness $d$ located at $z=0$. Combining the expression for the 3d current density of the vortex
\be\label{b2}
\j=\frac{c}{4\pi\lambda^2}\left(\frac{\Phi_0}{2\pi
r}\hat{\theta}-\A\right)\delta(z)d
\ee
where $d$ is the thickness, and the Maxwell's equation, we have
\be\label{b1}
\nabla^2\A=-\frac{4\pi}c\j=\frac
d{\lambda^2}\left(\A-\frac{\Phi_0}{2\pi
r}\hat{\theta}\right)\delta(z).
\ee
Next, we Fourier transform both sides of Eqn. (\ref{b1}):
\begin{equation}\label{8}
-\A(\q,k_z)=\frac1{(\q^2+k_z^2)}\frac{d}{\lambda}^2\left(\A(\q,z=0)-\frac{\Phi_0}{iq}\hat{\theta_q}\right),
\end{equation}
where $\q$ is the $2d$ wave vector, $k_z$ is the wave vector in $z-$direction, and $\hat{\theta_q}$ is the azimuthal unit vector in $q-$space. Defining the inverse 2d screening length $q_c=d/(2\lambda^2)$ and
integrating both sides $\int_{-\infty}^{\infty} \dd k_z$, one obtains
\begin{equation}\label{9}
\A(\q,z=0)=\frac{q_c}{q+q_c}\frac{\Phi_0}{iq}\hat{\theta_q}.
\end{equation}
From (\ref{b2}), we have
\be\label{b_j}
\j(\q)=\frac{q_c}{q+q_c}\frac{c{\Phi_0}}{2\pi i}\hat{\theta_q}.
\ee
Now, we calculate the interaction potential between two vortices in a single superconducting thin film. The first vortex is located at $r=0$, whose current distribution is given by (\ref{b_j}):
\be\label{j1}
\j_1(\q)=\frac{q_c}{q+q_c}\frac{c{\Phi_0}}{2\pi i}\hat{\theta_q}.
\ee
The second one is located at $\vec R$ away from the origin:
\begin{equation}\label{j2}
\begin{aligned}
\j_2(\q)&=\int \dd^2r\j_2(\r)e^{-i\q\cdot\r}=\int
\dd^2r\j_1(\r+\R)e^{-i\q\cdot\r}\\
&=\j_1(\q)e^{i\q\cdot\R}.
\end{aligned}
\end{equation}
Their interaction potential is given by
\be\label{U_single_layer}
U(\vec R)=\frac{2\pi}{c^2}\int\frac{\dd^2q}{(2\pi)^2}\left(\frac1{q_c}+\frac1q\right)\j_1(-\q)\j_2(\q),
\ee
where the first term is the kinetic energy contribution, while the second the term is from the magnetic energy $B^2$ term. Using (\ref{j1}) and (\ref{j2}), we have
\be
\begin{aligned}
U(\vec R)&=\frac{2\pi}{c^2}\int\frac{\dd^2q}{(2\pi)^2}\left(\frac1{q_c}+\frac1q\right)\j_1(-\q)\j_1(\q)e^{i\q\cdot\R}\\
&=\int\frac{\dd^2q}{(2\pi)^2}\frac{\Phi_0^2q_c}{2\pi}\frac1{q(q+q_c)}e^{i\q\cdot\R}\\
&\equiv\int\frac{\dd^2q}{(2\pi)^2}U(q)e^{i\q\cdot\R},
\end{aligned}
\ee
where the vortex interaction potential
\be
U(q)=\frac{\Phi_0^2q_c}{2\pi}\frac1{q(q+q_c)}
\ee
is exactly the same as what we obtained earlier in Appendix \ref{duality} with field theory formalism.

For the case of bilayer thin films with interlayer separation $a$, we can proceed in the same way. But there is one subtlety in that case. A vortex in layer 1, characterized by a phase singularity in layer 1, will also induce a circulating screening current in layer 2. Suppose the two identical layers are located at $z=0$ and $z=-a$, respectively, the one-vortex configuration is given by
\be
\begin{aligned}\label{j1j1'}
&\j_{1}=\frac{c}{4\pi\lambda^2}\left(\frac{\Phi_0}{2\pi
r}\hat{\theta}-\A(z=0)\right)\delta(z)d,\\
&\j_{1}'=\frac{c}{4\pi\lambda^2}\left(-\A(z=-a)\right)\delta(z+a)d,\\
&\nabla^2\A=-\frac{4\pi}c\left(\j_{1}+\j_{1}'\right).
\end{aligned}
\ee
Performing Fourier transform, one obtains
\be
\begin{aligned}
\A(\q,k_z)&=\frac{2q_c}{q^2+k_z^2}\\
&\times\left(\frac{\Phi_0}{iq}\hat{\theta_q}-\A(\q,z=0)-e^{ik_za}\A(\q,z=-a)\right)\nonumber.
\end{aligned}
\ee
Integrating over $k_z$, one obtains two equations for $\A(\q,z=0)$ and $\A(\q,z=-a)$, whose solution is given by
\be
\begin{aligned}
\A(\q,z=0)&=\frac{q_c[q+q_c(1-e^{-2qa})]}{(q+q_c)^2-q_c^2e^{-2qa}}\times\frac{\Phi_0}{iq}\hat{\theta_q},\\
\A(\q,z=-a)&=\frac{q_cqe^{-qa}}{(q+q_c)^2-q_c^2e^{-2qa}}\times\frac{\Phi_0}{iq}\hat{\theta_q}.
\end{aligned}
\ee
Thus, one can obtain $\j_1$ and $\j_1'$ from (\ref{j1j1'})
\be
\begin{aligned}
\j_1&=\frac{q_c(q+q_c)}{(q+q_c)^2-q_c^2e^{-2qa}}\frac{c\Phi_0}{2\pi i}\hat{\theta_q},\\
\j_1'&=-\frac{q_c^2e^{-qa}}{(q+q_c)^2-q_c^2e^{-2qa}}\frac{c\Phi_0}{2\pi i}\hat{\theta_q}.
\end{aligned}
\ee

\begin{figure}
\includegraphics[scale=0.35]{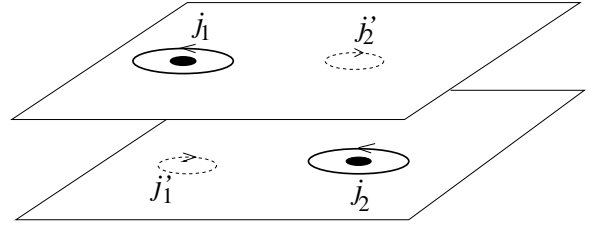}
\caption{The setup for calculating vortex interlayer interaction potential $U_e$. A phase singularity in layer 1 leads to current $\j_1$ and $\j_1'$ in laye 1 and 2, respectively, and similarly a phase singularity in layer 2 leads to current $\j_2$ and $\j_2'$ in layer 2 and 1, respectively. }\label{pearl}
\end{figure}

Next, one put in the currents $\j_2$ and $\j_2'$ of another vortex either in the same layer or the other layer, and calculate the intralayer and interlayer vortex interaction potential $U_i$ and $U_e$ in the same way as we did for the single layer case. For example, to calculate the vortex interlayer interaction $U_e$, we put in another vortex with its core at the second layer, and it has a current $\j_2$ in the second layer, and a circulating screening current $\j_2'$ in the first layer (see FIG. \ref{pearl}). Thus,
\be
\begin{aligned}
U_e(\vec R)&=\frac{2\pi}{c^2}\int\frac{\dd^2q}{(2\pi)^2}\left[\left(\frac{1}{q_c}+\frac1q\right)(\j_1\j'_2+\j_2\j_1')\right.\\
&+\left.\frac{e^{-qa}}q(\j_1\j_2+\j_1'\j_2')\right].
\end{aligned}
\ee
The final results are exactly the same as what we found in the field theory formalism in Sec. \ref{sec:bilayer} and Appendix \ref{app:bilayer}:
\be
\begin{aligned}
U_i(q)
&=\frac{\Phi_0^2q_c}{2\pi}\frac{q+q_c}{q(q^2+2q_cq+q_c^2(1-e^{-2qa}))},\\
U_e(q)&=-\frac{q_c}{q+q_c}e^{-qa}U_i.
\end{aligned}
\ee


\section{Classical hard-disc liquid description of the vortex metal phase}\label{appendix:hard-disc}

As explained in Sec. \ref{sec:metal}, we expect that our results for
the vortex drag do not depend sensitively on the microscopic model we
use for the vortices. In Sec. \ref{sec:metal} we used the fermionic
vortex response function to determine the drag resistance in the
intermediate metallic regime. Here we demonstrate the robustness of
this result by reproducing the drag resistance results while modeling
the vortex liquid in this regime as a classical hard-disc liquid.

The density response function $\chi(k,z)$ for a liquid of hard-core disks in the hydrodynamical limit is\cite{Leutheusser1982,Leutheusser1983,Forster}
\be\label{eqn:hd}
\chi(k,z)=\chi(k)+i\frac zT C(k,z),
\ee
where $z$ is the frequency, $T$ is the temperature, $\chi(k)$ is the static compressibility, and
\be
\begin{aligned}
C(k,z)&=iT\chi(k)\left[\frac1{\gamma}\frac{z+ik^2(\Gamma+D(\gamma-1))}{z^2-c^2k^2+izk^2\Gamma}\right.\\
&\left.+\left(1-\frac1{\gamma}\right)\frac1{z+ik^2D}\right],
\end{aligned}
\ee
showing a diffusive mode with weight $1-\frac1{\gamma}$, and a propagating mode with velocity $c$, weight $1/\gamma$ and life time $1/(\Gamma k^2)$.
Thus
\be
\frac{\chi(k,z)}{\chi(k)}=\left(1-\frac1{\gamma}\right)\frac{Dk^2}{Dk^2-iz}+\frac1{\gamma}\frac{c^2k^2-izDk^2(\gamma-1)}{c^2k^2-z^2-i\Gamma k^2z},
\ee
which satisfies the defining property of $\chi$:
\be
\chi(k)=\lim_{z\rightarrow0}\chi(k,z).
\ee
Here, $\gamma={C_p}/{C_v}$,
$C_v=1$ is the constant volume specific heat, and
\be
C_p=C_v+{T\chi_T\beta_V^2}/n
\ee
is the constant pressure specific heat, where $n$ is the vortex density, $\chi_T=\frac1{nT}\lim_{k\rightarrow0}S(k)$ is the isothermal compressibility, and $S(k)$ is the structure factor of the vortex liquid; $\beta_V\equiv n(1+y)$, where $y\equiv\frac{\pi}2n\sigma^2g(\sigma)$, 
\be
g(\sigma)
\equiv\frac{1-7\zeta/16}{(1-\zeta)^2}-\frac{\zeta^3/64}{(1-\zeta)^4},
\ee
$\zeta=\frac{\pi n\sigma^2}4$ is the packing fraction, and $\sigma$ is the diameter of the hard-disc vortex which we take to be the core size of the vortex, which in turn is approximately superconducting coherence length $\sim10$nm.\\

In addition, $\Gamma = a \left(\frac{\gamma-1}{\gamma}\right)+b$, and the diffusion coefficient $D=\frac a{\gamma}$, where 
\be
\begin{aligned}
a&=\frac{\nu\sigma^2}4+\frac2{\nu}(1+3y/4)^2v_0^2,\\
b&=3\nu\sigma^2/8+v_0^2(1+y/2)^2/\nu,
\end{aligned}
\ee
 $\nu=2\sqrt{\pi}n\sigma g(\sigma)v_0$ is called the Enskog collision
frequency, and the thermal velocity $v_0=\sqrt{\frac{T}m}$, $m$ is the
vortex mass. Finally, the speed of sound is
\be
c=\sqrt{\frac{C_p}{C_v}}\frac{v_0}{nT\chi_T}.
\ee

The static compressibility $\chi(k)$ is related to the structure factor $S(k)$ (strictly speakly, the Ursell function \cite{Chaikin}) by
\be
\chi(k)=\frac{n}{T}S(k),
\ee
and the structure factor $S(k)$ of a hard disk liquid is determined by following the so-called Percus-Yevick approximation of Ref. \onlinecite{Leutheusser1986,Whitlock2007}:
\be 
S(k)=1/(1-n h(k)),
\ee
where
\be 
h(k)=2\pi\int_0^{\infty}\dd RRJ_0(kR)h(R),
\ee
\be 
h(R)=\left\{\begin{array}{cc}h(0)+\frac{\zeta h(1)^2S(R)}{2\mu_D}, & 0\leq R<1 \\ 0, & R\geq 1 \end{array}\right..
\ee
Here, $\mu_D=\pi/16$, $\zeta=\frac{\pi n\sigma^2}4$ is the packing fraction, 
\be 
h(1) = \frac{\sqrt{(1-4\zeta)^2-4(\alpha-\beta)}-(1-4\zeta)}{2(\alpha-\beta)},
\ee
\be 
h(0) = h(1) - \beta h(1)^2,
\ee
\be 
\beta=\frac{\zeta S(R=1)}{2\mu_D},\qquad \alpha=2\zeta^2 A,
\ee
\be 
A=\frac1{\mu_D}\left(\frac{2}{\tilde a}\right)^3\int_0^{\frac{\tilde a}2}\dd z z^2(1-z^2)^{1/2},\tilde a=1+\zeta,
\ee
\be 
S(R)=\frac1{\tilde a}\left\{\arcsin\left(\frac{\tilde a R}2\right)+\frac{\tilde a R}2\left[1-\left(\frac{\tilde a R}2\right)^2\right]^{1/2}\right\}.
\ee
\begin{figure}
\includegraphics[scale=0.45]{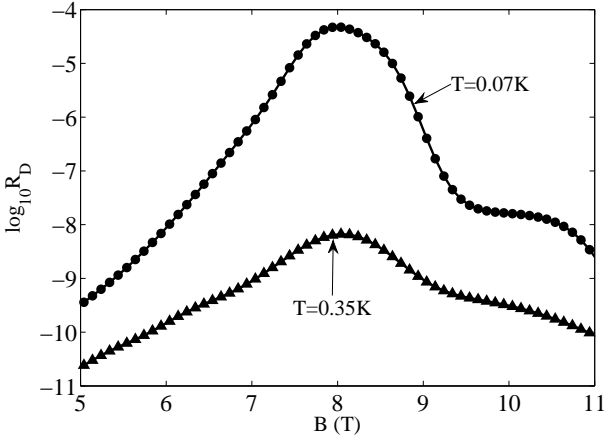}
\caption{Drag resistance in the vortex paradigm at $T=0.07$K, with the metallic phase modeled as classical hard-disc liquid. Everything else is the same those in FIG. 2.}\label{fig:hd}
\end{figure}

Putting these formulae together, we can compute the vortex density
response function in (\ref{eqn:hd}) and insert it into the drag
resistance formula (\ref{fermionic_vortex_drag}). The drag resistance
is shown in FIG. \ref{fig:hd}. One can see that it is remarkably close
to our results obtained in Sec. (\ref{sec:metal}), and thereby
demonstrating that the scale of the drag resistance in the metallic
regime is mainly set by the factors $dR/dB$ and is not sensitive to
the statistics of the vortex particles.

\section{Coulomb Drag for disordered electron glass}\label{Coulomb_drag}

In this section, we calculate the drag resistance due to Coulomb interaction between two disordered electron glasses with finite thickness. This calculation is related to the work of Ref. \onlinecite{Shimshoni1997}, but in our case the screening of the interlayer Coulomb interaction is important (see below), and we take into account the effect of finite film thickness.

The general formula for Coulomb drag resistance in $d$ dimensions is\cite{Zheng,Oreg}
\be\label{c1}
\rho_D^{ij}=\frac{\hbar^2}{e^2}\frac1{2\pi n^2T}\frac1{\Omega}
\sum_{\k}k^ik^j\int_0^{\infty}\frac{\dd\omega}{\sinh\frac{\hbar\omega}{2T}}|U|^2\im\chi_1\im\chi_2.
\ee
For the quasi-2d film we are considering, we can break the wavevector
summation into two summations: one over $k_z$, another over the 2d
wavevector $\q$. The $k_z$ summation is dominated by the term with
$k_z=0$ component, which physically corresponds to the configuration
with constant density along $z$-direction. In this case, we can use the quasi-2d form of the intralyer and interlayer Coulomb interaction potentials
\be
U_i(\q,k_z=0)=\frac{2\pi e^2d}{q},\qquad  U_e(\q,k_z=0)=\frac{2\pi e^2d}{q}e^{-qa},\nonumber
\ee
where $d$ is the film thickness, and $a$ is the center-to-center layer
separation. The real and imaginary parts of the density response
function for a localized electron gas is\cite{VollhardtWolfle1980,Vollhardt1992,Imry1982}
\begin{align}
\re\chi(\q,k_z=0,\omega)&=\nu (q^2+k_z^2)\xi^2\Big|_{k_z=0}=\nu q^2\xi^2, \nonumber\\
\im\chi(\q,k_z=0,\omega)&=\nu\frac{(q^2+k_z^2)\omega\xi^4}{D}\Big|_{k_z=0}=\nu\frac{q^2\omega\xi^4}{D},\nonumber
\end{align}
where $\nu$ is the 3d density of states at the Fermi energy, and $\xi$ is the localization length, and $D$ is the diffusion constant in the conducting phase. The above expression is valid so long as $\im\chi\ll\re\chi$, which is straightforward to verify in our case recalling that $\omega$ is cut off by the temperature $T$ in (\ref{c1}).

Thus, in the screened interlayer interaction we can neglect $\im\chi$ compared to $\re\chi$:
\begin{align}
U&=\frac{U_ie^{-qa}}{(1+U_i\chi_1)(1+U_i\chi_2)-(U_ie^{-qa}\chi_1)(U_ie^{-qa}\chi_2)}\nonumber\\
&\approx\frac1{2U_i\re\chi_1\re\chi_2\sinh(qa)},
\end{align}
where in the last line we have made an approximation that $U_i\re\chi\gg1$, i.e.,
\be
qa\gg x_0\equiv\frac{a}{\nu\xi^22\pi e^2d}.
\ee
We have verified that the contribution from $0<qa<x_0$ is negligible compared to that from $qa>x_0$.
Therefore,
\begin{align}
R_D&=\frac{\rho_D^{xx}}d\nonumber\\
&=\frac1{8\pi^2(nd)^2T}\frac{\hbar^2}{e^2}\int_{x_0}^{\infty}q^3\dd q\nonumber\\
&\times\int_0^{\infty}\frac{\dd\omega}{\sinh^2\frac{\hbar\omega}{2T}}\frac{\im\chi_1\im\chi_2}{4U_i^2(\re\chi_1)^2(\re\chi_2)^2\sinh^2(qa)}\nonumber\\
&=\frac{T^2}{128\pi^4\hbar e^2(nda)^2(D_1e^2d\nu)(D_2e^2d\nu)}\nonumber\\
&\times\int_{x_0}^{\infty}\frac{x\dd x}{\sinh^2x}\int_0^{\infty}\frac{x^2\dd x}{\sinh^2(x/2)}\nonumber\\
&=\frac{T^2}{128\pi^4\hbar e^2(nda)^2(D_1e^2d\nu)(D_2e^2d\nu)}\log\frac1{2x_0}\frac{4\pi^2}3\nonumber\\
&=\frac{T^2}{96\pi^2\hbar e^2(nda)^2(D_1e^2d\nu)(D_2e^2d\nu)}\log\frac1{2x_0}.\nonumber
\end{align}
Note that
\be
De^2d\nu=\frac1R,
\ee
we have
\begin{align}
R_D&=\frac{T^2R_1R_2}{96\pi^2\hbar e^2 (nda)^2}\log\frac1{2x_0}\nonumber\\
&=\frac1{96\pi^2}\frac{R_1R_2}{\hbar/e^2}\left(\frac{T}{e^2nda}\right)^2\log\frac1{2x_0}.
\end{align}
Since $D$ is the diffusion constant in the conducting phase, $R$ in
the above expression should also be the resistance of the conducting
phase. Thus this expression gives a slight overestimate of the drag
resistance in the percolation paradigm if we use the value of $R_{NN}$ of
the insulating phase for simplicity.

Note that our derivation relied on momentum summations. There are
concerns that such an approach, although quite common in the
literature, is incorrect when attempting to describe drag in strongly
disordered systems. For our purposes, the derivation based on
Eq. \ref{c1} is sufficient; this issue is taken up, however, in
Ref. \onlinecite{Apalkov2005}.

\section{The absence of measurable drag effect associated with a genuine superconductor in a resistor network}\label{SS_drag}

In this section, we show that a genuine superconducting link (i.e.,
without mobile vortices) has no measurable drag effect in a resistor
network.

\begin{figure}[h]
\includegraphics[scale=0.35]{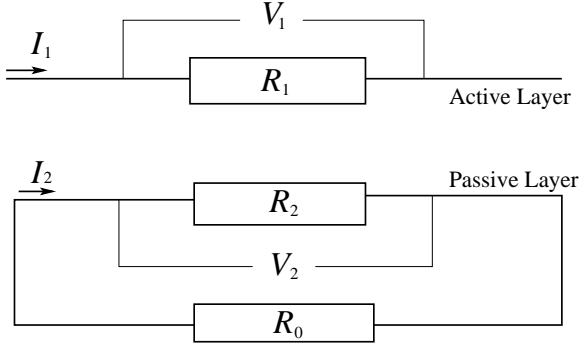}
\caption{The typical setup for a drag effect experiment: in the active
  layer, a driving current $I_1$ flows through a resistor $R_1$
  (normal or superconducting) with a voltage drop $V_1=I_1R_1$. In the
  passive layer, certain interaction effect takes place in a resistor
  $R_2$ (normal or superconducting), which may result in a drag
  current $I_2$ and a voltage drop $V_2$ across $R_2$. $R_2$ is also
  connected to another resistor $R_0$, which can be of any
  value.}\label{drag}
\end{figure}

FIG. \ref{drag} illustrates the typical setup for a drag effect
experiment: in the active layer, a driving current $I_1$ flows through
a resistor $R_1$ (normal or superconducting) with a voltage drop
$V_1=I_1R_1$. In the passive layer, certain interaction effects take
place in a resistor $R_2$ (normal or superconducting), which may
result in a drag current $I_2$ and a voltage drop $V_2$ across
$R_2$. $R_2$ is also connected to another resistor $R_0$, which might
represent a voltmeter, an open circuit ($R_0=\infty$), or something
else.

When one talks about the drag effect, there are two different concepts
one needs to distinguish. The first one is the "intrinsic" effect,
which manifests itself by the appearance of a drag current $I_D$ in
the passive layer {\it if $R_0=0$}. Generically, we have
\be
I_D\equiv I_2|_{R_0=0}=\eta I_1.
\ee
For example, for the case of $R_1,R_2>0$, i.e., both $R_1$ and $R_2$
are non-superconducting, $I_2|_{R_0=0}=\sigma_DV_1=\sigma_DI_1R_1$
(e.g., Coulomg drag between two 2DEGs), thus $\eta=\sigma_DR_1$; for
$R_1=R_2=0$ (superconductor), we have the Cooper pair version of the
supercurrent drag effect Eqn. (\ref{jj_drag}), thus $\eta$ is finite
in this case as well. For the case of $R_1>0$ (normal) and $R_2=0$
(superconducting), it would be unphysical to have $\eta=\infty$, thus
we have $\eta<\infty$ and $\sigma_{D,NS}=\eta/R_1<\infty$. From Kubo
formula for the drag conductance, we expect that
$\sigma_{D,SN}=\sigma_{D,NS}<\infty$, and hence for the case of
$R_1=0$ and $R_2>0$ we have $\eta=\sigma_{D,SN}R_1=0$.

In contrast, the second drag effect is the drag current $I_2$ in the presence of $R_0$, in which case he drag current at $R_0=0$ may or may not survive.
In a large-size resistor network we are considering for the
percolation picture, when we focus on the drag effect of one specific
link $R_2$, we can simplify the circuit of the passive layer to be of
the form in FIG. \ref{drag}, in which case $R_0$ representing the rest
of the circuit is almost always larger than $0$. If the drag effect
survives the presence of the nonzero $R_0$, it will manifest itself as
the appearance of a non-zero drag emf $\varepsilon_D$ on $R_2$. To see
this, first consider the case $R_2>0$, and $R_1$ can be either $0$ or
$>0$. $I_2$ receives contribution from both Ohm's law and the drag
effect:
\be
I_2=\frac{V_2}{R_2}+\eta I_1=-\frac{I_2R_0}{R_2}+\eta I_1,
\ee
thus
\be\label{normal}
I_2=\frac{(\eta R_2)I_1}{R_0+R_2}\equiv\frac{R_DI_1}{R_0+R_2}\equiv\frac{\varepsilon_D}{R_0+R_2},
\ee
where $\varepsilon_D=R_DI_1$ is the drag emf, and $R_D=\eta R_2$ is the drag resistance. If $R_1=0$ (superconducting) and $R_2>0$ (normal), we argued earlier that $\eta=0$, and thus $\varepsilon_D=R_D=0$ and there is no drag effect.

If $R_2=0$ (superconductor), no matter if $R_1=0$ (superconducting) or
$>0$ (normal), it is straightforward to see from Kirchoff's Law that
we have only one steady-state solution $I_2|_{R_0>0}=0$. More insight
into this case can be gained by considering what happens in real
time. Suppose at time $t=0$, the drag effect takes place, a drag
supercurrent $I_2(R_0=0)$ starts to flow in the circuit. But due to
the presence of the normal resistor $R_0$, a voltage $I_2R_0$ now
exist on the supercondutor, which will crank up the phase winding of
the superconductor and degrade the drag supercurrent, until a steady
state is reached where the total supercurrent is zero. Thus, we see
that for the case $R_2=0$ and $R_0>0$, there is no observable drag
effect, i.e., $I_2|_{R_0>0}=0$,  $\varepsilon_D=I_2(R_2+R_0)=0$,
$R_D=\varepsilon_D/I_1=0$, although there is nonzero ``intrinsic''
drag effect $\eta$.

We can also understand this result $R_D=0$ for $R_2=0$ by examining
the expression $R_D=\eta R_2$. For both the case of $R_1=R_2=0$ and
the case of $R_1>0$ and $R_2=0$, we found earlier that $\eta<\infty$,
and thus the drag resistance $R_D=\eta R_2$ and the drag emf
$\varepsilon_D$ are $0$ for $R_2=0$.

In conclusion, we have shown that when connected with a nonzero
resistor, as typically true in a resistor network, a genuine
superconducting link has no measurable drag effect at all, no matter
whether it is vertically aligned with a normal link or another
superconducting link.

\bibliography{reference}

\end{document}